\documentclass[ twoside,openright,titlepage,fleqn,numbers=noenddot,headinclude,
                11pt,a4paper,BCOR5mm,footinclude=true,cleardoublepage=empty,abstractoff 
                ]{scrreprt}
\newcommand{\myTitle}{Polar codes in quantum information theory\xspace}
\newcommand{\myDegree}{Master's thesis\xspace}
\newcommand{\myName}{Christoph Hirche\xspace}
\newcommand{\myProf}{Prof. Dr. Reinhard F. Werner\xspace}
\newcommand{\myOtherProf}{Prof. Dr. Tobias J. Osborne\xspace}
\newcommand{\mySupervisor}{Dr. Ciara Morgan\xspace}

\newcommand{\myTime}{January 6, 2015\xspace}
\newcommand{\myBirthdate}{September 03, 1990\xspace}
\newcommand{\myBirthplace}{Hannover\xspace}

\usepackage[latin1]{inputenc} 
\usepackage[square,numbers]{natbib} 
\usepackage[fleqn]{amsmath} 
\usepackage{float} 
\usepackage{classicthesis-ldpkg} 
\usepackage[eulerchapternumbers,listings,
            subfig,beramono,eulermath,parts,dottedtoc,floatperchapter]{classicthesis}

\usepackage{makerobust} 
\makeatletter 
\MakeRobustCommand\caption@xref 
\makeatother 

\usepackage[english]{iktthesis}
\hypersetup{%
    colorlinks=true, linktocpage=true,%
    breaklinks=true, pageanchor=true,%
    plainpages=false, bookmarksnumbered, bookmarksopen=true, bookmarksopenlevel=1,%
    hypertexnames=true, urlcolor=webbrown, linkcolor=RoyalBlue, citecolor=webgreen,%
}

\usepackage[english]{babel}
\usepackage{quantum}  
\usepackage{tikz}
\usepackage{amsmath}

\usetikzlibrary{}
\usepgfmodule{oo}

\allowdisplaybreaks[1]
\numberwithin{equation}{section}

\usepackage[nottoc]{tocbibind} 
\usepackage[scaled]{helvet}
\usepackage{url}

\begin{document}
\frenchspacing
\raggedbottom
\pagenumbering{roman}
\pagestyle{plain}
 
\pgfdeclarelayer{background}
\pgfdeclarelayer{firstbackground}
\pgfdeclarelayer{secondbackground}
\pgfsetlayers{secondbackground,firstbackground,background,main}

\pdfbookmark[0]{Titelblatt}{title}
\begin{titlepage}
  \changetext{}{19mm}{}{19mm}{}
  \vspace{1cm}
  \begin{center}
    \includegraphics[width=13.8cm]{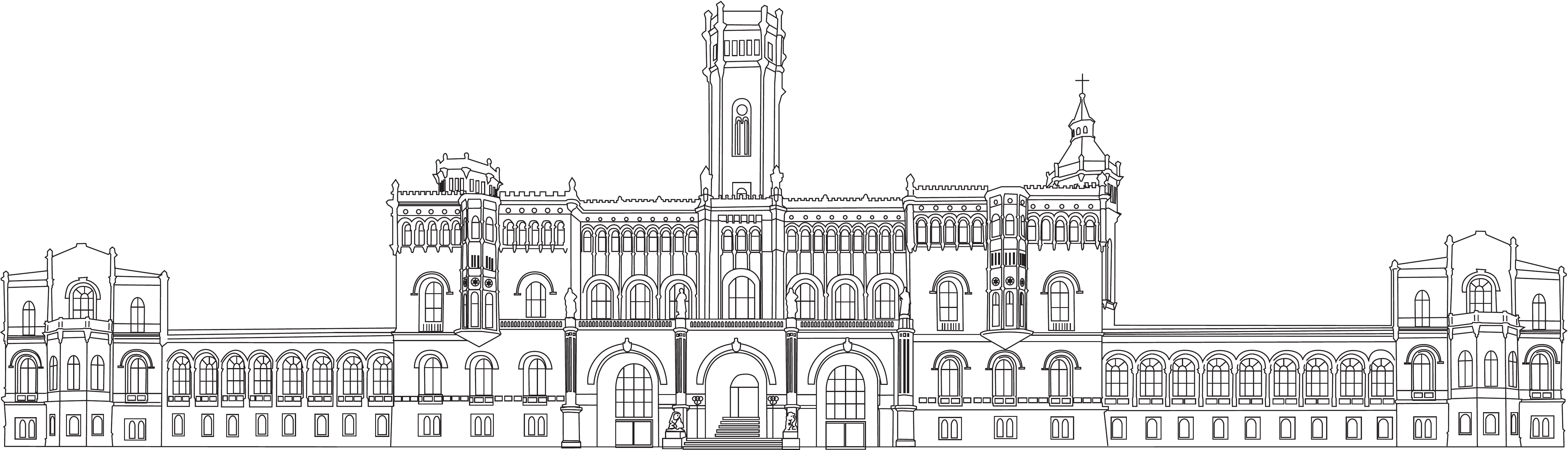} \\ 
  \end{center}
  \medskip
  \begin{center}
    \textbf{\huge\spacedallcaps{L}\LARGE\spacedallcaps{eibniz}}
    \textbf{\huge\spacedallcaps{U}\LARGE\spacedallcaps{niversit\"{a}t}}
    \textbf{\huge\spacedallcaps{H}\LARGE\spacedallcaps{annover}} \\
  \end{center}

  \begin{center}
    \normalsize
    \spacedallcaps{Fakult\"{a}t}
    \spacedallcaps{f\"{u}r}
    \spacedallcaps{Mathematik}
    \spacedallcaps{und}
    \spacedallcaps{Physik} \\
    \smallskip
    \spacedallcaps{Institut}
    \spacedallcaps{f\"{u}r}
    \spacedallcaps{theoretische}
    \spacedallcaps{Physik}
  \end{center}
  \vfill
  \vfill

  \condWIWI{
  \begin{center}
    \Large \myTitle
  \end{center}
  \vfill
  \vfill

  \begin{center}
    \LARGE \textbf{\myDegree}
  \end{center}
  \vfill

  \begin{center}
    \large zur Erlangung des Grades eines Diplom-Wirtschaftsingenieurs der \\
    \large Fakult\"{a}t f\"{u}r Elektrotechnik und Informatik, Fakult\"{a}t f\"{u}r Maschinenbau und \\
    \large Wirtschaftswissenschaftlichen Fakult\"{a}t der Leibniz Universit\"{a}t Hannover
  \end{center}
  \vfill

  \begin{center}
    \large vorgelegt von \\
  \end{center}
  \medskip

  \begin{center}
    \large \spacedallcaps{\myName}
  \end{center}
  \medskip

  \begin{center}
    \large geb. am \myBirthdate in \myBirthplace \\
  \end{center}

  \vfill
  \vfill

  Pr\"{u}fer: \myProf
  \bigskip

  Hannover, den \myTime
  } 
  {
  \begin{center}
    \LARGE \myTitle
  \end{center} 
  \vfill
  \vfill

  \begin{center}
    \LARGE \textbf{\myDegree}
  \end{center}
  \vfill

  \begin{center}
    \large \spacedallcaps{\myName}
  \end{center}

  \begin{center}
    \large \myTime \\
  \end{center} 
  \vfill

  \begin{center}
    \begin{tabular}{lll}
      1st supervisor  & : & \myProf \\
     2nd supervisor & : & \myOtherProf \\
      co-supervisor   & : & \mySupervisor
    \end{tabular}
  \end{center} 

  }  



  \changetext{}{-19mm}{}{-19mm}{}

\end{titlepage}

\thispagestyle{empty}

\hfill

\vfill

\noindent\myName: \textit{\myTitle,} \myDegree, \textcopyright\ \myTime

%
%
%
%
%

\cleardoublepage
\pdfbookmark[1]{Abstract}{Abstract}
\begingroup
\let\clearpage\relax
\let\cleardoublepage\relax
\let\cleardoublepage\relax

\chapter*{Abstract}
Polar codes are the first capacity achieving and efficiently implementable codes for classical communication. Recently they have also been generalized to communication over classical-quantum and quantum channels. In this work we present our recent results for polar coding in quantum information theory, including applications to classical-quantum multiple access channels, interference channels and compound communication settings, including the first proof of channel coding achieving the Han-Kobayashi rate region of the interference channel without the need  of a simultaneous decoder. 

Moreover we add to the existing framework by extending polar codes to achieve the asymmetric capacity and improving the block error probability for classical-quantum channels. In addition we use polar codes to prove a new achievable rate region for the classical-quantum broadcast channel. 

We also discuss polar codes for quantum communication over quantum channels and state results towards codes for compound quantum channels in this setting. 

We conclude by stating a list of interesting open questions to invite further research on the topic.

\vfill

\endgroup			

\vfill

\cleardoublepage
\cleardoublepage
\pdfbookmark[1]{Abstract}{Abstract}
\begingroup
\let\clearpage\relax
\let\cleardoublepage\relax
\let\cleardoublepage\relax

\chapter*{Acknowledgments}
Firstly I would like to thank Reinhard F. Werner and Tobias J. Osborne for taking over the supervision of my thesis. 

I am especially grateful to Ciara Morgan for the enormous amount of help, requiring a lot of time and work, in the process of writing this thesis. 

I am also very grateful to Mark M. Wilde for the enjoyable and fruitful collaboration, including introducing me to the topic. 

Last but not least I would like to thank all members of the Quantum Information Group in Hannover for making my time here so pleasant.

\vfill

\endgroup			

\vfill

\cleardoublepage

\pagestyle{scrheadings}

\cleardoublepage
\refstepcounter{dummy}
\pdfbookmark[1]{\contentsname}{tableofcontents}
\setcounter{tocdepth}{2} 
\setcounter{secnumdepth}{3} 
\manualmark
\markboth{\spacedlowsmallcaps{\contentsname}}{\spacedlowsmallcaps{\contentsname}}
\tableofcontents 

\cleardoublepage

\cleardoublepage
\automark[section]{chapter}
\renewcommand{\chaptermark}[1]{\markboth{\spacedlowsmallcaps{#1}}{\spacedlowsmallcaps{#1}}}
\renewcommand{\sectionmark}[1]{\markright{\thesection\enspace\spacedlowsmallcaps{#1}}}
\refstepcounter{dummy}
\pdfbookmark[1]{\listfigurename}{lof}
\listoffigures

\vspace*{8ex}

\cleardoublepage

\pagenumbering{arabic}

\chapter{Introduction}
One of the main tasks in classical and quantum information theory is to determine the optimal rate at which we can reliably send information from one party to another. These optimal rates are called the \emph{capacity} of a channel. To show that a certain rate is optimal requires two steps, first we have to show \emph{achievability}, that is that there exists a code to achieve the proposed rate. Then we need a \emph{converse} statement, showing that we can not do better. 

Besides the capacity of single-sender single-receiver channels we are also interested in settings with more parties involved, such as multiple access channel, broadcast channel and interference networks or settings with an uncertainty about the chosen communication channel, for example compound channel, where we do not know which channel will be used but only that it comes from a defined set of channels. 

Finding the achievable rate regions is very difficult in many cases. Indeed, for the classical interference channel we know the region only for the setting of very strong interference~\cite{Carleial75} or strong interference~\cite{HK81}. The best known achievable rate region for the classical two-user interference channel is given by the Han-Kobayashi rate region~\cite{HK81}. 
A quantum version of the Han-Kobayashi rate region for the classical-quantum interference channel
has also been conjectured to be achievable, based on the conjectured existence of a three-sender simultaneous decoder \cite{SFWSH11, FHSSW12}, and it was in fact proven to be achievable in \cite{S11} using a specialized
three-sender quantum simultaneous decoder.

Whether a quantum simultaneous decoder exists in general and also for settings with more then three senders remains unclear. Another approach to decoding the received information is successive decoding. Indeed, this has been successfully used in settings including the classical-quantum multiple access channel \cite{W01} and it would be desirable to extend successive decoding to other communication scenarios, such as the interference channel. This possibility was an open problem until now~\cite{FS12}. \\

Another important goal for information theory is not only to know the optimal rate, but also to specify a constructive way of coding to achieve the capacity. Along with communication rates one is also interested in obtaining a code that can be implemented efficiently, usually meaning with at most polynomial number of gates. \\

In the setting of classical inputs and outputs a new code was recently introduced, called "polar codes" \cite{A09}. Arikan showed that these codes can achieve the symmetric capacity of any classical single-sender single-receiver channel. Remarkably this can be done with a complexity $O(N\log N)$ for encoding and decoding, where $N$ is the number of channels used. 

Moreover, polar codes make use of the effect of channel polarization, where a recursive construction is used to divide the instances of a channel into a fraction that can be used for reliable communication and a fraction that is nearly useless. The crucial feature is that the fraction of \emph{good} channels is equal to the symmetric capacity of the channel. 

Polar codes attracted a lot of attention and were generalized to many additional settings, such as multiple access channels \cite{STY10, STY13, O13, MELK13}, broadcast channels \cite{GAG13, MHSU14}, interference networks \cite{WS14}
and for the task of source coding \cite{A10, A12} and universal coding for
compound channels \cite{HU13, MELK132, WS14}.

Recently polar codes have also been generalized to the setting of sending classical information over quantum channel \cite{WG13,WG12} as well as sending quantum information \cite{WR12, WG13QDeg, SRDR13}. For the task of sending classical information the quantum setting has also been generalized to multi-user settings, namely multiple access channel and interference channel as well as compound channels and compound multiple access channel by the author and collaborators in~\cite{HMW14}. 

It can also be shown in the quantum setting that the encoding can be done with the same efficiency as in the classical case~\cite{WG13,WG12,WR12}. The question of whether this is also possible for the decoding is in general left open \cite{WLH13}, but was positively answered for certain classes of channels, namely Pauli and erasure channels \cite{RDR12}. \\

In this work we will first review applications of polar codes in quantum information theory, based on \cite{WG13} for classical communication over quantum channels and generalize this scheme to achieve the asymmetric capacity and improve its block error probability. Also we present the results obtained by the author and collaborators in~\cite{HMW14} for multiple access channel, interference channel, compound channel and compound multiple access channel. Showing that the best known rate regions can be achieved using polar codes and a successive decoder, including the Han-Kobayashi rate region for the classical-quantum two-user interference channel.

We will also show how to apply polar codes to the classical-quantum broadcast channel to achieve a rate region larger then the ones previously known. Hence this is an example of how polar codes can be used to prove the achievability part of a channel coding theorem. 

Then we will discuss different approaches to polar codes for the transmission of quantum information over quantum channels \cite{WR12, SRDR13} and present possible extensions for universal polar codes for quantum communication over compound channels. \\

The thesis is organized as follows. In Section \ref{notation} we will state general preliminaries and in Section \ref{cqMulti} introduce the various types of channels discussed in this work. Chapter \ref{cqc} is about classical-quantum communication, in detail Section \ref{cq} describes the single-sender single-receiver channel, achieving the symmetric capacity in Section \ref{scq}, the asymmetric capacity in Section \ref{asy} and an improved block error probability in Section \ref{imp}. In Section \ref{cqUChannel} we discuss the compound channel, in Section \ref{MAC} the multiple access channel, in Section \ref{comMAC} the compound multiple access channel, in Section \ref{int} the interference channel and in Section  \ref{cqBroadcast} the broadcast channel. 
Chapter \ref{qqc} introduces quantum communication over quantum channels, where Section \ref{qChannel} describes the quantum point to point channel and Section \ref{UQC} the compound quantum channel. 
Finally in Chapter \ref{CO} we describe possible further directions for exploring polar codes in quantum communication settings in Section \ref{fd} and conclude with Section \ref{Conclusion}. 
\chapter{Preliminaries}
\section{Notation and definitions}\label{notation}

In this section we will introduce and explain all the necessary tools required for the remaining parts of the thesis. 

In the remaining work we will use the following notation for vectors: 
$u_1^N\equiv u^N$ denotes a row vector $(u_1, \dots, u_N)$ and correspondingly $u_i^j$ denotes, for $1\leq i, j\leq N$, a subvector $(u_i, \dots, u_j)$.
Note that if $j<i$ then $u_i^j$ is empty. Similarly, for a vector $u_1^N$ and
a set $A \subset \{ 1,\dots, N\}$, we write $u_A$ to denote the subvector $%
(u_i : i\in A)$.

We will consider quantum channels $\NN$, given by 
\begin{equation}
\NN : \LL(\HS_A) \rightarrow \LL(\HS_B),
\end{equation}
where $\LL(\HS)$ denotes the space of all operators on $\HS$. 
Due to the Stinespring dilation theorem~\cite{S55} we can also express a quantum channel in terms of an isometry $U_{BE} : \HS_A \rightarrow \HS_B\otimes\HS_E$ with an environment $\HS_E$ as 
\begin{equation}\label{stine}
\NN(\rho) = \tr_E (U_{BE}\rho U_{BE}^{\dagger})
\end{equation}
for every $\rho\in S_=(\HS_A)$, where $S_=(\HS_A)$ denotes the set of density operators on $\HS_A$. \\

A discrete classical-quantum channel $W$ takes realizations $x \in \mathcal{X%
}$ of a random variable $X$ to a quantum state, denoted $\rho_x^B$, on a
finite-dimensional Hilbert space $\mathcal{H}^B$, 
\begin{equation}
W : x \rightarrow \rho_{x}^{B},
\end{equation}
where each quantum state $\rho_x$ is described by a positive semi-definite
operator with unit trace. We will take the input alphabet $\mathcal{X} =
\{0,1\}$ unless otherwise stated, and the tensor product $W^{\otimes N}$ of $%
N$ channels is denoted by $W^N$.

To characterize the behavior of the classical-quantum channels we consider, we will make use of the
symmetric Holevo capacity, defined as follows: 
\begin{equation}
I(W) \equiv I(X;B)_\rho,
\end{equation}
where the quantum mutual information with respect to a classical-quantum
state $\rho^{XB}$ is given by  
\begin{equation}
I(X;B) \equiv H(X)_\rho + H(B)_\rho - H(XB)_\rho,
\end{equation}
with 
\begin{equation}\label{cqstate}
\rho^{XB} = \frac{1}{2} | 0\rangle\!\langle 0 | \otimes \rho_0^B + 
\frac{1}{2} | 1\rangle\!\langle 1 | \otimes \rho_1^B.
\end{equation}
In the above, the von Neumann entropy $H(\rho)$ is defined as 
\begin{equation}
H(\rho) \equiv -\tr\{\rho \log_2 \rho\}.
\end{equation}

A data-processing inequality holds for the mutual information~\cite{W13}. 
That is, suppose we have quantum states $\rho_{AC} \equiv \NN^{B\rightarrow C}(\sigma_{AB})$ then the following relation between the mutual information with respect to the states $\rho$ and $\sigma$ holds
\begin{equation} \label{dpi}
I(A;B)_{\sigma} \geq I(A;C)_{\rho}.
\end{equation}

We will also make use of the conditional von Neumann entropy defined as 
\begin{equation}
H(X\mid B)_\rho = H(X)_\rho - H(XB)_\rho
\end{equation}
and the quantum conditional mutual information defined
for a tripartite state $\rho^{XYB}$ as 
\begin{equation}
I(X;B\mid Y)_\rho \equiv H(XY)_\rho + H(YB)_\rho - H(Y)_\rho - H(XYB)_\rho.
\end{equation}
Operational interpretations for the conditional mutual information were found in~\cite{DY06} and recently in~\cite{FR14}. \\

We characterize the reliability of a channel $W$ as the fidelity between the
output states 
\begin{equation}
F(W) \equiv F(\rho_0, \rho_1),
\end{equation}
with 
\begin{equation}
F(\rho_0, \rho_1) \equiv \normTr{\sqrt{\rho_0}\sqrt{\rho_1}}^2
\end{equation}
and 
\begin{equation}
\normTr{A} \equiv \tr{\sqrt{A^\dagger A}}.
\end{equation}
Note that in the case of two commuting density matrices, $\rho = \sum_i p_i \ketbra{}{i}{i}$ and $\sigma = \sum_i q_i \ketbra{}{i}{i}$ the fidelity can be written as
\begin{equation}
F(\rho, \sigma) = \sum_i \sqrt{p_i q_i}.
\end{equation} 

The Holevo capacity and the fidelity can be seen as quantum generalizations
of the mutual information and the Bhattacharya parameter from the classical
setting, respectively (see, e.g., \cite{A09}).

We will also use the quantity 
\begin{equation}\label{Z}
Z(X\mid B)_{\rho} \equiv 2\sqrt{p_0p_1}\, F(\rho_0, \rho_1),
\end{equation}
for a classical-quantum state, as in Equation \ref{cqstate}, which can again be seen as quantum generalization of the Bhattacharya parameter, for a classical variable now with quantum side information. \\

For the transmission of quantum information over quantum channels, the capacity is described in terms of the coherent information defined as 
\begin{equation} 
I(A\rangle B)_\rho = - H(A\mid B)_\rho.
\end{equation}
The best known rate~\cite{L97, S02, D05} for a quantum channel $\NN^{A'\rightarrow B}$ is given by the regularized coherent information
\begin{equation}\label{qcapcity}
Q(\NN) = \lim_{n\rightarrow\infty}\frac{1}{n}Q^{(1)}(\NN^{\otimes n}),
\end{equation}
with
\begin{equation}\label{qcapcity}
Q^{(1)}(\NN) = \max_{\Phi_{AA'}\in S_=(AA')} I(A\rangle B)_{(\id\otimes\NN)\Phi_{AA'}}.
\end{equation}

\newpage
\section{Multi-user channels}

\label{cqMulti}

In the following sections, we will focus on two particular kinds of
multi-user channels: classical-quantum multiple access channels (cq-MACs)
and classical-quantum interference channels.

We begin with the classical-quantum interference channel, and for simplicity
we focus on the case of
two senders and two receivers.
The interference channel can
be modeled mathematically as the following triple: 
\begin{equation}
\left( \mathcal{X}_{1}\times \mathcal{X}_{2},W,\mathcal{H}^{B_{1}}\otimes 
\mathcal{H}^{B_{2}}\right) ,
\end{equation}%
with 
\begin{equation}
W:x_{1},x_{2}\rightarrow \rho _{x_{1},x_{2}}^{B_{1}B_{2}}.
\end{equation}
The information
processing task for the classical-quantum interference channel \cite%
{FHSSW12,SFWSH11} is as follows. The $k$th sender
would like to communicate a message to the $k$th
receiver, where $k \in \{1,2\}$. Sender $k$ chooses a message $m_k$ from a message set $\mathcal{M}%
_k = \{1, \cdots, 2^{nR_k} \}$, and encodes her message as a codeword $%
x_k^n(m_k) \in \mathcal{X}_k^n$.
The encoding for each sender is given by $%
\{x_{1}^{n}(m_{1})\}_{m_{1}\in \mathcal{M}_{1}}$ and $\{x_{2}^{n}(m_{2})%
\}_{m_{2}\in \mathcal{M}_{2}}$, respectively, with the corresponding
receivers' decoding POVMs denoted by $\{\Lambda _{m_{1}}\}$ and $\{\Gamma _{m_{2}}\}$.
The code is said to be a $(n,R_{1},R_{2},\epsilon )$-code, if the average
probability of error is bounded as follows 
\begin{equation}
\bar{p}_{e}=\frac{1}{|\mathcal{M}_{1}||\mathcal{M}_{2}|}%
\sum_{m_{1},m_{2}}p_{e}(m_{1},m_{2})\leq \epsilon ,
\end{equation}%
where the probability of error $p_{e}(m_1, m_2)$ for a pair of messages $(m_{1},m_{2})$
is given by 
\begin{equation}
p_{e}(m_{1},m_{2})=\mathrm{Tr}\left\{\left( I-\Lambda _{m_{1}}\otimes \Gamma
_{m_{2}}\right) \rho _{x_{1}^{n}(m_{1}),x_{2}^{n}(m_{2})}^{B^n_{1}B^n_{2}}\right\},
\end{equation}
with $\rho _{x_{1}^{n}(m_{1}),x_{2}^{n}(m_{2})}^{B^n_{1}B^n_{2}}$ the state resulting
when senders 1 and 2 transmit the codewords $x_{1}^{n}(m_{1})$
and $x_{2}^{n}(m_{2})$ through $n$ instances of the channel, respectively.

A rate pair $(R_1, R_2)$ is said to be \emph{achievable} for the two-user
classical-quantum interference channel described above if there exists an $%
(n, R_1,R_2, \epsilon)$-code $\forall \epsilon >0$ and large enough $n$. \\

The two-user cq-MAC is defined by the following triple corresponding to the
input alphabets, channel output state and output system 
\begin{equation}
\left( \mathcal{X}_{1}\times \mathcal{X}_{2},\rho _{x_{1},x_{2}}^{B},\mathcal{H}^{B}\right) .
\end{equation}
The coding task is for two senders to communicate individual messages to a
single receiver. The detailed description of the information processing task is
somewhat similar to the above, so we omit it for brevity's sake. \\

The two-user classical-quantum interference channel induces two c-q MACs
which can be modeled as 
\begin{equation}
\left( \mathcal{X}_{1}\times \mathcal{X}_{2},\rho _{x_{1},x_{2}}^{B_{1}}=%
\mathrm{Tr}_{B_{2}}\left\{\rho _{x_{1},x_{2}}^{B_{1}B_{2}} \right\},\mathcal{H}%
^{B_{1}}\right) ,
\end{equation}%
and 
\begin{equation}
\left( \mathcal{X}_{1}\times \mathcal{X}_{2},\rho _{x_{1},x_{2}}^{B_{2}}=%
\mathrm{Tr}_{B_{1}} \left\{\rho _{x_{1},x_{2}}^{B_{1}B_{2}}\right\},\mathcal{H}%
^{B_{2}}\right) .
\end{equation}

The \emph{rate region} for a channel is given by the closure of all
achievable rates for that channel. We will be particularly interested in the
Han-Kobayashi rate region for the two-user interference channel. This region
was achieved in the classical setting by exploiting a coding strategy for
the interference channel which induces two three-user MACs,
together with a simultaneous decoder~\cite{HK81}. \\

We will also investigate \emph{compound} channel. Indeed, compound channels form a class
of channels with so-called ``channel uncertainty.'' In this model, a channel
is chosen from a set of possible channels, and used to transmit the
information, thus generalizing the traditional setting in which both sender
and receiver have full knowledge of the channel before choosing their code.
The classical and quantum capacities of compound quantum channels have been
studied in \cite{BBN08, BB09}, respectively. 

A compound cq-channel is defined by a set $\mathcal{W} = \{ W_i\}$ of cq-channels where each $W_i$ can be written as
\begin{equation}
W_i : x \rightarrow \rho_{x, i}^{B},
\end{equation}
and characterized by its the output state $\rho_{x,i}^{B}$, taken with respect to the input $x \in \mathcal{X}$.

Later, we will be particularly interested in \emph{compound} cq-MACs, which are direct generalizations of compound channels  to the scenario of multiple access channels. 

A compound cq-MAC is defined by a set $\mathcal{W} = \{ W_i\}$ of cq-MAC channels where each $W_i$ can be written as
\begin{equation}
W_i : x_1 , x_2 \rightarrow \rho_{x_1,x_2,i}^{B},
\end{equation}
and characterized by its the output state $\rho_{x_1,x_2,i}^{B}$, taken with respect to the input pairs $(x_1,x_2) \in \mathcal{X}_1 \times \mathcal{X}_2$.

Note that for compound channels as well as for compound MACs we will look at the case in which the receiver knows the particular channel $W_i$ which has been chosen. However, the sender does not have this knowledge. This assumption can be easily justified in the case of taking many uses of the channel, since channel tomography can be performed in order to give the receiver knowledge of the channel which has been chosen. Moreover, this requires a small number of channel uses when compared to the overall number of channel uses, thereby not affecting the communication rate. 

We will also investigate two-user classical-quantum broadcast channels modeled by 
\begin{equation}
\left( \mathcal{X}_{1},W,\mathcal{H}^{B_{1}}\otimes 
\mathcal{H}^{B_{2}}\right) ,
\end{equation}
with 
\begin{equation}
W : x \rightarrow  \rho_{x}^{B_1B_2}.
\end{equation}


\newpage 
\chapter{Classical-Quantum Channel}\label{cqc}
In this chapter we present results on polar codes for several classical-quantum communication models, obtained by the author and collaborators, which can be found in \cite{HMW14}.
Moreover we will show how to achieve the asymmetric capacity of a channel in Section \ref{asy}, improve the block error probability in Section \ref{imp} and add polar codes for the classical-quantum broadcast channel to the framework in Section \ref{cqBroadcast}. 

\section{Polar codes for classical-quantum channels}\label{cq}

\subsection{Achieving the symmetric capacity}\label{scq}

The effect of polarization which will be applied to classical-quantum multi-user communication settings relies on polar codes for the single-user classical-quantum channel. A scheme achieving the capacity in this setting was introduced by Wilde and Guha in \cite{WG13}. In this section we will review their scheme in order to extend it in the following sections. 

Polar codes exploit the effect of channel polarization, which is achieved in
two steps, namely, by so-called channel combining and channel splitting.

\pgfooclass{stamp}{ 
    \method stamp() { 
    }
    \method apply(#1,#2,#3) { 
	\draw (#1+2,#2) -- (#1,#2) -- (#1,#2+1) -- (#1+2,#2+1) -- cycle;
          \node[font=\tiny] at (#1+1,#2+0.5) {#3};
    }
    \method box(#1,#2,#3,#4,#5) { 
	\filldraw[fill=#5] (#1+#3,#2) -- (#1,#2) -- (#1,#2+#4) -- (#1+#3,#2+#4) -- cycle;
   }
 \method cnot(#1,#2,#3,#4) { 
	\draw (#1,#2) -- (#1-#3,#2) -- (#1-#3,#2-#4) -- (#1,#2-#4)-- (#1-#3-2,#2-#4);
	\draw (#1-#3,#2) circle (0.2);
	\draw (#1-#3,#2+0.2) -- (#1-#3,#2) -- (#1-#3-0.25,#2);
	\draw (#1-#3-0.25,#2) -- (#1-#3-2,#2);
   }
}
\pgfoonew \mystamp=new stamp()

\pgfooclass{block}{ 
    \method block() { 
    }
    \method basic(#1,#2,#3) { 
	\mystamp.apply(#1,#2,#3)
	\draw (#1-1,#2+0.5) -- (#1,#2+0.5);
	\draw (#1+2,#2+0.5) -- (#1+3,#2+0.5);
    }
}

\pgfoonew \myblock=new block()

We start by building an intuition for the \textit{channel combining}  step, stating and illustrating the first three steps. 

For the first step we simply define $W_1$ as the chosen communication channel $W$.
\begin{figure}[htbp]
	\begin{minipage}[b]{0.4\textwidth} 
		\begin{equation*}
		W_1^{B_1} \equiv W : u_1 \rightarrow \rho_{u_1}^{B_1}
		\end{equation*}
	\end{minipage}
	\hfill
	\begin{minipage}[b]{0.4\textwidth}
\centering
 \begin{tikzpicture}[scale=0.5]
	 \myblock.basic(0, 0, $W$)
        \node[font=\tiny] at (1,1.5) {$W_1$};
\end{tikzpicture}
	\end{minipage}
\end{figure}

For the second step we combine two $W_1$ channels by a CNOT gate to get a channel, denoted $W_2$: 
\begin{figure}[htbp]
	\begin{minipage}[b]{0.4\textwidth} 
\begin{equation*}
W_2^{B_1B_2} : (u_1, u_2) \rightarrow \rho^{B_1}_{u_1\oplus u_2} \otimes \rho^{B_1}_{u_2}
\end{equation*}
	\end{minipage}
	\hfill
	\begin{minipage}[b]{0.4\textwidth}
\centering
\begin{tikzpicture}[scale=0.5]
	\myblock.basic(0, 0, $W_1$)
	\myblock.basic(0, 2, $W_1$)
	\mystamp.cnot(0,2.5,1,2)
    \begin{pgfonlayer}{secondbackground}
	\mystamp.box(-1.5,-0.5,4,4,blue!3);
    \end{pgfonlayer}
        \node[font=\tiny] at (0.5,4) {$W_2$};
\end{tikzpicture}
	\end{minipage}
\end{figure}

Then for the third step we again combine two $W_2$ channel by CNOT gates and swap two bits in order to combine two inputs of the same type of $W_2$.

This gives us a channel, denoted $W_4$: 
\begin{figure}[h!]
	\begin{minipage}[b]{0.6\textwidth} 
\begin{align*}
&W_4^{B_1B_2B_3B_4} : (u_1, u_2,u_3,u_4) \\ 
&\rightarrow W_2^{B_1B_2}(u_1\oplus u_2, u_3\oplus u_4) \otimes W_2^{B_3B_4}(u_2, u_4)
\end{align*}
	\end{minipage}
	\hfill
	\begin{minipage}[h]{0.3\textwidth}
\centering
\begin{tikzpicture}[scale=0.45]
	\draw (0,0) -- (1,0) -- (1,2) -- (0,2) -- cycle;
          \node[font=\tiny] at (0.5,1) {$W_2$};
	\draw (0,3) -- (1,3) -- (1,5) -- (0,5) -- cycle;
          \node[font=\tiny] at (0.5,4) {$W_2$};
	\mystamp.cnot(-2,1.5,1,1)
	\mystamp.cnot(-2,4.5,1,1)
	\draw (-2,1.5) -- (-1,3.5) -- (0,3.5);
	\draw (-2,3.5) -- (-1,1.5) -- (0,1.5);
	\draw (-2,4.5) -- (0,4.5);
	\draw (-2,0.5) -- (0,0.5);
	\draw (1,4.5) -- (2.5,4.5);
	\draw (1,3.5) -- (2.5,3.5);
	\draw (1,1.5) -- (2.5,1.5);
	\draw (1,0.5) -- (2.5,0.5);
          \node[font=\tiny] at (-1,6) {$W_4$};
    \begin{pgfonlayer}{secondbackground}
	\mystamp.box(-3.5,-0.5,5,6,blue!3);
    \end{pgfonlayer}
    \begin{pgfonlayer}{firstbackground}
	\mystamp.box(-2.5,0,2,5,blue!5);
    \end{pgfonlayer}
\end{tikzpicture}
	\end{minipage}
\end{figure}

Iterating this procedure we can build arbitrary $W_{2^n}$. \\

An alternative way to describe the channel $W_N$, induced by $N$ single copies of the channel $W$, is by a linear transformation, 
given by $x^N=u^NG_N$ transforming the input sequence $u^N$, where 
\begin{equation}
G_N = B_NF^{\otimes n}
\end{equation}
with 
\begin{equation}
F \equiv \left[ 
\begin{matrix}
1 & 0 \\ 
1 & 1
\end{matrix}
\right],
\end{equation}
and $B_N$ is a permutation matrix known as a ``bit reversal'' operation \cite{A09}. \\

The second step is \textit{channel splitting}, where the combined channel $W_N$ from the
previous step is used to define new channels $W^{(i)}_N$ as follows:
\begin{equation}
W^{(i)}_N : u_i \rightarrow \rho_{(i),u_i}^{U_1^{i-1}B^N},
\end{equation}
where 
\begin{equation}
\rho^{U^{i-1}_1 B^N}_{(i),u_i} = \sum_{u_1^{i-1}} \frac{1}{2^{i-1}} %
\ketbra{}{u_1^{i-1}}{u_1^{i-1}} \otimes \sum_{u_{i+1}^N} \frac{1}{2^{N-i}}
\rho^{B^N}_{u^N}.
\end{equation}

This is equivalent to a decoder which estimates, by the $i$-th measurement, the bit $u_i$, with the following
assumptions: the entire output is available to the decoder, the previous
bits $u_1^{i-1}$ are correctly decoded and the distribution over the bits $u^N_{i+1}$ is uniform.
The assumptions that all previous bits are correctly decoded is called ``genie-aided'' and can be ensured by a limited amount of classical communication prior to the information transmission. 
The decoder described above is thus a ``genie-aided'' successive cancelation decoder. 

It can easily be seen that with each iteration step of the polar the capacity of code some split channels become higher and others become lower, while the overall sum of capacities stay the same. 

Thus these two steps give rise to the effect of channel polarization, which ensures that the fraction of channels $%
W_N^{(i)}$ which have the property $I(W_N^{(i)}) \in (1-\delta, 1]$ goes to
the symmetric Holevo information $I(W)$ and the fraction with $I(W_N^{(i)})
\in [0, 1-\delta)$ goes to $1-I(W)$ for any $\delta \in (0,1)$, as $N$ goes to
infinity through powers of two. This is one of the main insights of the work by Arikan \cite{A09} and the generalization in \cite{WG13} (see \cite{WG13} for a more detailed statement). 

Hence we choose a polar code as a ``$G_N$%
-coset code'' \cite{A09}; that is, we choose a subset $A \subset
\{1,\dots,N\}$ and re-write the input transformation 
\begin{equation}
x^N=u^NG_N
\end{equation}
as 
\begin{equation}
x^N=u_AG_N(A) \oplus u_{A^c} G_N(A^c),
\end{equation}
where $G_N(A)$ denotes the submatrix of $G_N$ constructed from the rows of $%
G_N$ with indices in $A$. 
Now we can fix a code 
\begin{equation}
(N,K,A,u_{A^c})
\end{equation}
 where $N$ is the length of the code, $A$
fixes the indices for the information bits, $K=|A|$ is the number of information bits and $u_{A^c}$ is the vector of
so-called ``frozen bits''. The sender and receiver need to exchange the values of these frozen bits before hand, giving rise to the need of classical communication before starting the protocol. 

A polar code has the above properties and is such that it obeys the
following rule, that is the set of information indices $A$ is
chosen such that the following inequality between the fidelities of the $i$th and $j$th split channel
\begin{equation}
F(W^{(i)}_N) \leq F(W^{(j)}_N)
\end{equation}
holds for all $i\in A$ and $j\in A^c$.
This is called the polar coding rule. 

Lastly, a bound on the block error probability $P_e(N,R)$ for blocklength $N$ and rate $R$ was derived for a fixed $%
R<I(W)$ and $\beta< \frac{1}{2}$, with the result as shown in \cite{WG13} 
\begin{equation}
P_e(N,R) = o(2^{-\frac{1}{2}N^\beta}).
\end{equation}
The measurement by which this error bound is achieved was called a quantum successive
cancellation decoder \cite{WG12}, and the error analysis mainly exploited Sen's non-commutative union bound \cite{S11}.

\newpage 

\subsection{Achieving the asymmetric capacity}\label{asy}

For asymmetric channels the capacity is not always achieved by an uniform input distribution.
Several approaches to shaping the input distribution to achieve the asymmetric capacity are known for communication over classical channels \cite{HY13, G68, SRDR12}. 
In this section we show how to achieve the asymmetric capacity for classical-quantum channels, generalizing an approach for classical channels presented in \cite{HY13}. 

To achieve this goal we divide the polarization process into two consecutive steps. Before we actually consider the channel, we need to shape our given uniform input distribution such that it is optimal for the chosen communication channel. This task is precisely the reverse task of lossless compression of a given source with non-uniform input distribution. Since this step is completely classical, also for the classical-quantum channel, we can simply consider classical lossless compression, a task for which a polar coding approach is described in \cite{K09}. 

In this setting, an input message $X_1^N$ with non-uniform distribution is transformed into $U_1^N = X_1^N G_N$, with $G_N$ being the usual polar coding transform. It is shown in \cite{K09} that $U_1^N$ polarizes in the sense that a fraction of bits is uniformly distributed and independent of all previous bits, we call this set $\FF$, and the remaining bits can be completely determined by their previous bits, we call this set $\FF^c$. 

For our application we need the reverse protocol. This is done as follows, first we determine the sets $\FF$ and $\FF^c$ for the desired distribution. Then we fill $U_1^N$ with uniformly distributed bits for $i\in\FF$ and with bits determined from $U_1^{i-1}$ for $i\in\FF^c$. Now we can simply transform $U_1^N$ to $X_1^N$ by $X_1^N=U_1^N G_N^{-1}=U_1^N G_N$, where the last equality follows from the fact that the polar coding transform is its own inverse. 
Thus, given a mostly uniformly distributed $U_1^N$ we can get a $X_1^N$ with a shaped distribution such that it achieves the optimal rate for the used asymmetric channel. 

Form \cite{HY13} we know that the size of the polarized sets is as follows: 
\begin{align}
\lim_{N\rightarrow\infty} \frac{1}{N} | \FF | &= H(X) \\
\lim_{N\rightarrow\infty} \frac{1}{N} | \FF^c | &= 1-H(X).
\end{align}

In the second step we consider the channel with classical input $X_1^N$ and quantum output $B^N$, which now corresponds to lossless source coding with quantum side information. This means that the set $\FF$ polarizes further into a  set $\II$, where it follows from $H(U_i\mid U_1^{i-1}, B^N) \leq H(U_i\mid U_1^{i-1})$, since conditioning does not increase entropy,  that $\II\subset\FF$,  and a set $\FF\backslash\II$. where $\II$ denotes the set of indices from $U_1^N$ which are uniformly distributed but can be determined from $U_1^{i-1}$ and $B^N$, these will be the information bits. While $\FF\backslash\II$ denotes those which are independent of $U_1^{i-1}$ and $B^N$ and as such will have to be set as frozen bits. 

This effect relies on the polarization of $Z(U_i \mid U_1^{i-1}, B_1^N)$, which is defined by Equation \ref{Z} and is a direct generalization of the classical conditional Bhattacharya parameter, now with quantum side information. The fact that this quantity polarizes is shown in Appendix C of \cite{SRDR13}, where it is also shown that this quantity polarizes simultaneously with the conditional von Neumann entropy. 
Hence we know that the sets polarize as follows: 
\begin{align}
\lim_{N\rightarrow\infty} \frac{1}{N} \left| \left\{ i\in  [N] : Z(U_i\mid U_1^{i-1},B^N) \leq 2^{-N^{\beta}}\right\}  \right| &= 1 - H(X\mid B) \\
\lim_{N\rightarrow\infty} \frac{1}{N} \left| \left\{ i\in  [N] : Z(U_i\mid U_1^{i-1},B^N)  \geq 2^{-N^{\beta}}\right\} \right| &= H(X\mid B).
\end{align}

It now remains to determine the size of the set $\II$,
\begin{align}
\lim_{N\rightarrow\infty} \frac{1}{N} | \FF\backslash\II \cup \FF^c | &=& \lim_{N\rightarrow\infty} \frac{1}{N} \left( | \FF\backslash\II| +  |\FF^c | \right) &=& 1 - I(X;B) \\
\lim_{N\rightarrow\infty} \frac{1}{N} | \II\cap\FF | &=& 1 - \lim_{N\rightarrow\infty} \frac{1}{N} | \FF\backslash\II \cup \FF^c | &=& I(X;B).
\end{align}

Hence we can code with the rate given by the mutual information maximized over possible input distributions and therefore achieve the asymmetric capacity of the channel, known as the Holevo capacity
\begin{equation}
C(W) = \max_{p(x)} I(X;B) . 
\end{equation}

\newpage

\subsection{Improved block error probability} \label{imp}

In this section we will show how a technique recently developed by Gao in \cite{J14}, can be used to improve the block error probability for classical-quantum polar codes. 

As discussed in Section \ref{scq} the block error probability was shown in \cite{WG13} to be 
\begin{equation}
P_e(N,R) = o(2^{-\frac{1}{2}N^\beta}).
\end{equation}
This was derived using Sen's non-commutative union bound \cite{S11}, given by 
\begin{equation} 
1 - \tr\{\Pi_N \dots \Pi_1\rho \Pi_1\dots\Pi_N\} \leq 2 \sqrt{\sum_{i=1}^N \tr\{(\id - \Pi_i)\rho\}}. 
\end{equation}
Recently Gao improved this bound \cite{J14} by showing that the following inequality also holds
\begin{equation} 
1 - \tr\{\Pi_N \dots \Pi_1\rho \Pi_1\dots\Pi_N\} \leq 4 \sum_{i=1}^N \tr\{(\id - \Pi_i)\rho\}. 
\end{equation}
We can apply this bound in place of Sen's in the calculation of the block error probability for classical-quantum polar codes to also improve this bound to 
\begin{equation}
P_e(N,R) = o(2^{-N^\beta}).
\end{equation}

In the following we will show how to derive this bound. The calculation goes along the lines of that of the error probability in \cite{WG13}. As in \cite{WG13} we start with the definition of the error probability and then apply Gao's bound: 
\begin{align*}
&P_e(N,K,A) \\
&= \frac{1}{2^N} \sum_{u^N} \left( 1-\tr\{ \Pi^{B^N}_{(N),u_1^{N-1}u_N} \dots \Pi^{B^N}_{(1),u_1} \rho_{u^N} \Pi^{B^N}_{(1),u_1} \dots \Pi^{B^N}_{(N),u_1^{N-1}u_N} \} \right) \\
&\leq \frac{1}{2^N} \sum_{u^N} 4 \sum_{i=1}^N \tr\{\left( \id - \Pi^{B^N}_{(i),u_1^{i-1}u_i}\right) \rho_{u^N} \}
\end{align*}
where the $\Pi^{B^N}_{(i),u_1^{i-1}u_i}$ denote the measurements which determine whether the $i$th input was zero or one (for details see the explicit construction in \cite{WG13} or the generalization to the MAC in Section \ref{2MAC}). \\

Continuing, to bound $P_e(N,K,A)$, 
\begin{align*}
&= \frac{1}{2^N} \sum_{u^N} 4 \sum_{i\in A} \tr\{\left( \id - \Pi^{B^N}_{(i),u_1^{i-1}u_i}\right) \rho_{u^N} \} \\
&=  4 \sum_{u^N} \sum_{i\in A} \frac{1}{2^N}\tr\{\hat\Pi^{B^N}_{(i),u_1^{i-1}u_i} \rho_{u^N} \} \\ 
&=  4  \sum_{i\in A}\sum_{u_i} \frac{1}{2}\sum_{u^{i-1}_1} \frac{1}{2^{i-1}}  \tr\left\{\hat\Pi^{B^N}_{(i),u_1^{i-1}u_i} \sum_{u^{N}_{i+1}} \frac{1}{2^{N-i}}\rho_{u^N} \right\}
\end{align*}
where the first equality follows from the fact that we can simply set measurements for frozen bits to be the identity and for the second equality we define
\begin{equation*}
\hat\Pi^{B^N}_{(i),u_1^{i-1}u_i} = \id - \Pi^{B^N}_{(i),u_1^{i-1}u_i}.
\end{equation*}
The third equality simply follows from the expansion and reordering of the sum and normalization. 

Continuing the upper bound on the block error probability $P_e(N,K,A)$, 
\begin{align*}
&=  4  \sum_{i\in A} \sum_{u_i} \frac{1}{2}\sum_{u^{i-1}_1} \frac{1}{2^{i-1}} \tr\left\{\hat\Pi^{B^N}_{(i),u_1^{i-1}u_i} \bar\rho_{u^N} \right\} \\
&=  4  \sum_{i\in A} \sum_{u_i} \frac{1}{2} \tr\left\{ \left( \id - \sum_{u^{i-1}_1} \ketbra{}{u_1^{i-1}}{u_1^{i-1}}^{U^{i-1}_1} \otimes \Pi^{B^N}_{(i),u_1^{i-1}u_i} \right) \right.\\
 &\hphantom{=  4  \sum_{i\in A} \sum_{u_i} \frac{1}{2} \tr\id - \Pi^{B^N}_{(i),u_1^{i-1}u_i}}\left.\sum_{u^{i-1}_1}\frac{1}{2^{i-1}} \ketbra{}{u_1^{i-1}}{u_1^{i-1}}^{U^{i-1}_1} \otimes \bar\rho_{u^N} \right\} 
\end{align*}
where the first equality follows by defining the average state
\begin{equation*}
\bar\rho_{u^N} = \sum_{u^{N}_{i+1}} \frac{1}{2^{N-i}}\rho_{u^N}
\end{equation*}
and the second equality from the fact \cite{WG13} that
\begin{align*}
\sum_x p(x)\tr\{A_x\rho_x\} = \tr\left\{ \left( \sum_x \ketbra{}{x}{x} \otimes A_x \right) \left( \sum_{x'} p(x')\ketbra{}{x'}{x'} \otimes \rho_{x'} \right) \right\}. 
\end{align*}

Continuing the bound on the block error probability $P_e(N,K,A)$, 
\begin{align*}
&=  4  \sum_{i\in A} \sum_{u_i} \frac{1}{2} \tr\left\{\left( \id - \Pi^{U_1^{i-1}B^N}_{(i),u_i} \right) \rho^{U_1^{i-1}B^N}_{(i),u_i} \right\}  \\
&\leq 4  \sum_{i\in A} \frac{1}{2} \sqrt{F(W^{(i)})}.
\end{align*}
The equality follows from the definition of $\Pi^{U_1^{i-1}B^N}_{(i),u_i}$ in \cite{WG13}, these can be seen as measurements with additional classical information containing all previous bits. 
The inequality follows, as in \cite{WG13}, using Lemma 3.2 from \cite{H06}, which states an upper bound on the minimum of the sum of two error probabilities. 

Finally using Theorem 3 in \cite{WG13} there exists a sequence of sets $A_N$ with size $|A_N| \geq NR$ for any $R<I(W)$ and $\beta<\frac{1}{2}$ such that 
\begin{align}
\sum_{i\in A_N} \sqrt{F(W^{(i)})} = o(2^{-N^{\beta}}), 
\end{align}
and thus 
\begin{align}
P_e(N,K,A) \leq 4 \sum_{i\in A}  \frac{1}{2} \sqrt{F(W^{(i)})} = o(2^{-N^{\beta}}),
\end{align}
which completes the proof.

\newpage
\section{Universal polar codes for cq-channel}\label{cqUChannel}

\pgfdeclarelayer{background}
\pgfdeclarelayer{firstbackground}
\pgfdeclarelayer{secondbackground}
\pgfsetlayers{secondbackground,firstbackground,background,main}

\pgfooclass{pol}{ 
    \method pol() { 
    }
    \method apply(#1,#2) { 
	\draw (#1+0.5,#2) -- (#1,#2) -- (#1,#2+4) -- (#1+0.5,#2+4) -- cycle;
    }
    \method box(#1,#2,#3,#4,#5) { 
	\filldraw[fill=#5] (#1+#3,#2) -- (#1,#2) -- (#1,#2+#4) -- (#1+#3,#2+#4) -- cycle;
   }
 \method cnot(#1,#2,#3,#4) { 
	\draw (#1,#2) -- (#1-#3,#2) -- (#1-#3,#2-#4) -- (#1,#2-#4);
	\draw (#1-#3,#2) circle (0.1);
	\draw (#1-#3,#2+0.1) -- (#1-#3,#2) -- (#1-#3-0.25,#2);
	\draw[dotted] (#1-#3-0.25,#2) -- (#1-#3-1.25,#2);
   }
}
\pgfoonew \mypol=new pol()

\pgfooclass{bblock}{ 
    \method bblock() { 
    }
    \method basic(#1,#2) { 
	 \mypol.apply(#1+2.5, #2+0.5)
           \mypol.apply(#1+2.5, #2+5)
    \begin{pgfonlayer}{background}
     	\mypol.box(#1, #2, 3.5, 9.5, black!15)
    \end{pgfonlayer}
	\mypol.cnot(#1+2.5,#2+8.5,1.5,7)
	\mypol.cnot(#1+2.5,#2+8,1.75,7)
    }
}

\pgfoonew \mybblock=new bblock()

\pgfooclass{step}{ 
    \method step() { 
    }
    \method first(#1,#2) { 
	\mybblock.basic(#1+2,#2+10.5)
	\mybblock.basic(#1+2,#2+0.5)
    \begin{pgfonlayer}{firstbackground}
	\mypol.box(#1+0,#2+0,6,20.5, black!10)
    \end{pgfonlayer}
	\mypol.cnot(#1+4.5,#2+14.5,3.5,8)
	\mypol.cnot(#1+4.5,#2+14,3.75,8)
    }
}

\pgfoonew \mystep=new step()

\pgfooclass{arrows}{ 
    \method arrows() { 
    }
    \method apply2() { 
          \draw[->] (3.5,19.5)node[font=\tiny, below right=-1.5]{1} -- (3.5,19); 			
	\draw[->,densely dotted] (3.5,19)node[font=\tiny,below right=-1.5]{6} -- (3.5,18.5);
	\draw[->] (3.5,18.5)node[font=\tiny, below right=-1.5]{7} -- (3.5,15.5);
          \draw[->] (3.5,15)node[font=\tiny, below right=-1.5]{1} -- (3.5,14.5); 			
	\draw[->,densely dotted] (3.5,14.5)node[font=\tiny, below right=-1.5]{4} -- (3.5,14);
	\draw[->] (3.5,14)node[font=\tiny, below right=-1.5]{5} -- (3.5,12);
          \draw[->,densely dotted] (3.5,12)node[font=\tiny, below right=-1.5]{6} -- (3.5,11.5);
	\draw[->] (3.5,11.5)node[font=\tiny, below right=-1.5]{7} -- (3.5,11);
          \draw[->] (3.5,9.5)node[font=\tiny, below right=-1.5]{1} -- (3.5,9);  			
	\draw[->,densely dotted] (3.5,9)node[font=\tiny, below right=-1.5]{2} -- (3.5,8.5);
	\draw[->] (3.5,8.5)node[font=\tiny, below right=-1.5]{3} -- (3.5,6.5);
          \draw[->,densely dotted] (3.5,6.5)node[font=\tiny, below right=-1.5]{4} -- (3.5,6);
	\draw[->] (3.5,6)node[font=\tiny, below right=-1.5]{7} -- (3.5,5.5);
	\draw[->] (3.5,5)node[font=\tiny, below right=-1.5]{1} -- (3.5,2);				
          \draw[->,densely dotted] (3.5,2)node[font=\tiny, below right=-1.5]{2} -- (3.5,1.5);
	\draw[->] (3.5,1.5)node[font=\tiny, below right=-1.5]{7} -- (3.5,1);
	\node[font=\tiny] at (4.5+0.25,1+2) {N};
	\node[font=\tiny] at (4.5+0.25,5.5+2) {N};
	\node[font=\tiny] at (4.5+0.25,11+2) {N};
	\node[font=\tiny] at (4.5+0.25,15.5+2) {N};
    }
    \method apply3() { 
          \draw[->] (3.5,9.5)node[font=\small, below right=0]{1} -- (3.5,9);  			
	\draw[->,densely dotted] (3.5,9)node[font=\small, below right=0]{2} -- (3.5,8.5);
	\draw[->] (3.5,8.5)node[font=\small, below right=0]{3} -- (3.5,7.5);
	\draw[->] (3.5,5)node[font=\small, below right=0]{1} -- (3.5,2);				
          \draw[->,densely dotted] (3.5,2)node[font=\small, below right=0]{2} -- (3.5,1.5);
	\draw[->] (3.5,1.5)node[font=\small, below right=0]{3} -- (3.5,1);
	\node[font=\tiny] at (4.5+0.25,1+2) {N};
	\node[font=\tiny] at (4.5+0.25,5.5+2) {N};
    }
}

\pgfoonew \myarrows=new arrows()

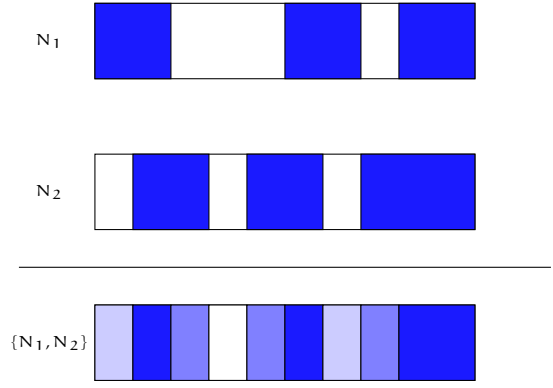
\begin{figure}[tbp]
\centering
\begin{tikzpicture}[scale=1]
	\draw (0,5) -- (5,5) -- (5,6) -- (0,6) -- cycle;
	\draw (0,3) -- (5,3) -- (5,4) -- (0,4) -- cycle;
	\draw (-1,2.5) -- (6,2.5);
	\draw (0,1) -- (5,1) -- (5,2) -- (0,2) -- cycle;
	\filldraw[fill=blue!90] (0,5) -- (1,5) -- (1,6) -- (0,6)  -- cycle;
	\filldraw[fill=blue!90] (2.5,5) -- (3.5,5) -- (3.5,6) -- (2.5,6)  -- cycle;
	\filldraw[fill=blue!90] (4,5) -- (5,5) -- (5,6) -- (4,6)  -- cycle;
	\filldraw[fill=blue!90] (0.5,3) -- (1.5,3) -- (1.5,4) -- (0.5,4)  -- cycle;
	\filldraw[fill=blue!90] (2,3) -- (3,3) -- (3,4) -- (2,4)  -- cycle;
	\filldraw[fill=blue!90] (3.5,3) -- (5,3) -- (5,4) -- (3.5,4)  -- cycle;
	\filldraw[fill=blue!20] (0,1) -- (0.5,1) -- (0.5,2) -- (0,2)  -- cycle;
	\filldraw[fill=blue!90] (0.5,1) -- (1,1) -- (1,2) -- (0.5,2)  -- cycle;
	\filldraw[fill=blue!50] (1,1) -- (1.5,1) -- (1.5,2) -- (1,2)  -- cycle;
	\filldraw[fill=white] (1.5,1) -- (2,1) -- (2,2) -- (1.5,2)  -- cycle;
	\filldraw[fill=blue!50] (2,1) -- (2.5,1) -- (2.5,2) -- (2,2)  -- cycle;
	\filldraw[fill=blue!90] (2.5,1) -- (3,1) -- (3,2) -- (2.5,2)  -- cycle;
	\filldraw[fill=blue!20] (3,1) -- (3.5,1) -- (3.5,2) -- (3,2)  -- cycle;
	\filldraw[fill=blue!50] (3.5,1) -- (4,1) -- (4,2) -- (3.5,2)  -- cycle;
	\filldraw[fill=blue!90] (4,1) -- (5,1) -- (5,2) -- (4,2)  -- cycle;
         \node[font=\tiny] at (-0.6,5.5) {$N_1$};
         \node[font=\tiny] at (-0.6,3.5) {$N_2$};
         \node[font=\tiny] at (-0.6,1.5) {$\{N_1, N_2\}$};
\end{tikzpicture}
\caption[Polar codes for different channels.]{Polar codes for different channels. We look at two channels $N_1$ and $N_2$ and mark the good indices in blue. For the compound channel $\{N_1, N_2\}$ based on the sets of these two channels some indices will be good for both channels (blue), good for just one channel (lighter shades of blue) or for none (white).}\label{compound}
\end{figure}

In this section we explore the polar coding technique for the classical-quantum compound channel, as defined in Section \ref{cqMulti}. We will describe how to apply the second scheme described in \cite{HU13} for classical compound channels to classical quantum channels.  
For our purposes we assume that the sender does not know which channel will be used for communication but the receiver does. 
This is a reasonable assumption, since the sender could simply send a limited amount of information through the channel to process channel tomography, such that the receiver can determine which channel was used. 
For the polar code construction we observe that the sets of good indicies depend highly on the particular channel. 
Therefore if we apply the encoder to a compound channel we will get indices which are good for one channel but not necessarily for the other one (see Figure \ref{compound}). 
For now we look at compound channels based on a set of two different channels $W$ and $V$. 
It is clear that we immediately know how to code for those indices which are good for both or bad for both channels.
The essential idea is to align indices which are only good for one channel with those which are only good for the other one. 
Let
\begin{equation}
\begin{split}\label{GGBB}
\mathcal{G}_{(1)}& =\left\{i\in \lbrack 1:N]:\sqrt{F(W_N^{(i)})}<2^{-N^{\beta }}\right\}, \\
\mathcal{G}_{(2)}& =\left\{i\in \lbrack 1:N]:\sqrt{F(V_N^{(i)})}<2^{-N^{\beta }}\right\}, \\
\mathcal{B}_{(1)}& =\left\{i\in \lbrack 1:N]:\sqrt{F(W_N^{(i)})}\geq 2^{-N^{\beta }}\right\}, \\
\mathcal{B}_{(2)}& =\left\{i\in \lbrack 1:N]:\sqrt{F(V_N^{(i)})}\geq 2^{-N^{\beta }}\right\}.
\end{split}%
\end{equation}%
denote the sets of indices corresponding to whether a bit is good or bad for a channel. 
Due to the polarization effect for single channels, all bits are guaranteed to be in one of the following sets: 
\begin{align}
\begin{aligned}\label{AAAA}
\mathcal{A}_{\text{I}}\ & =\mathcal{G}_{(1)}\cap \mathcal{G}_{(2)}, \\
\mathcal{A}_{\text{II}}\ & =\mathcal{G}_{(1)}\cap \mathcal{B}_{(2)}, \\
\mathcal{A}_{\text{III}}& =\mathcal{B}_{(1)}\cap \mathcal{G}_{(2)}, \\
\mathcal{A}_{\text{IV}}& =\mathcal{B}_{(1)}\cap \mathcal{B}_{(2)}.
\end{aligned}
\end{align}%
Bits corresponding to indices in the set  $\mathcal{A}_{\text{I}}$ will also be decoded with high probability in the compound setting and bits in $\mathcal{A}_{\text{IV}}$ will have to be fixed as frozen bits. 
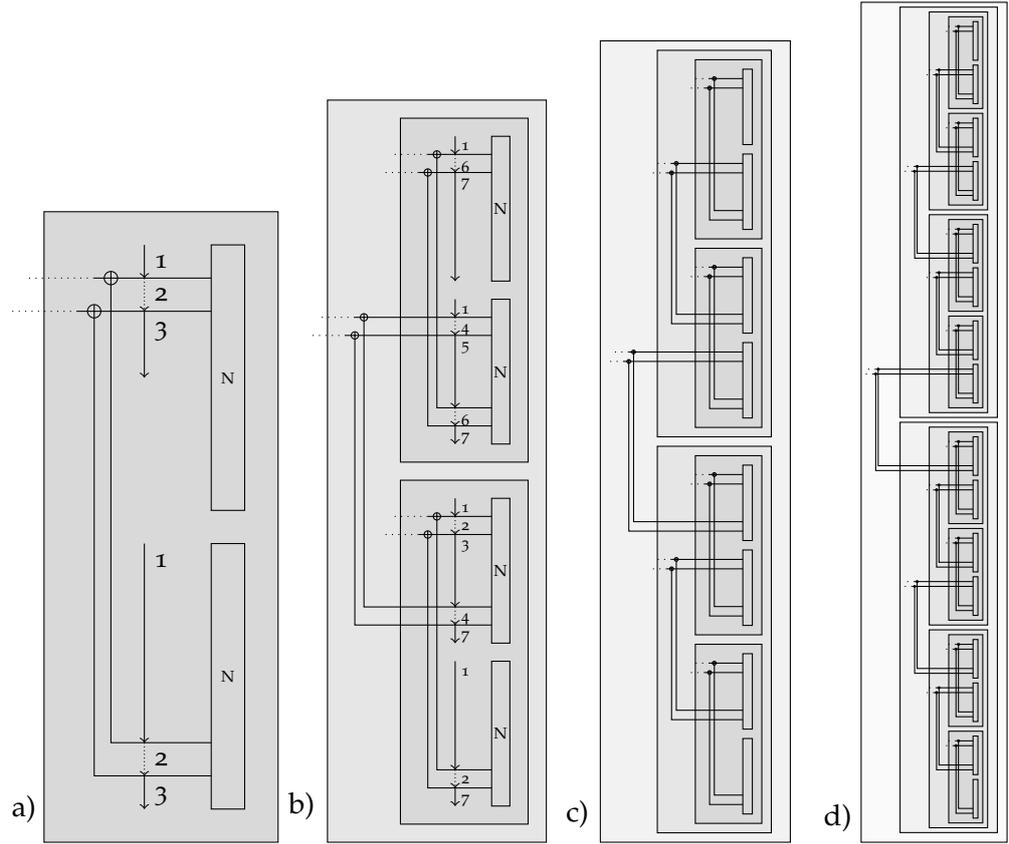
\begin{figure}[t!]
	\begin{minipage}[b]{0.24\textwidth} 
\begin{tikzpicture}[scale=0.88]
	\mybblock.basic(2,0.5)
	\myarrows.apply3()
		\node at (1.7,1) {a)};
\end{tikzpicture}
	\end{minipage}
	\hfill
	\begin{minipage}[b]{0.24\textwidth}
\begin{tikzpicture}[scale=0.48]
	\mystep.first(0,0)
	\myarrows.apply2()
\node at (-0.7,1) {b)};
\end{tikzpicture}
	\end{minipage}
	\hfill
	\begin{minipage}[b]{0.22\textwidth}
\begin{tikzpicture}[scale=0.25]
\node at (-4.2,1) {c)};
	\mystep.first(0,0)
	\mystep.first(0,21)
	\mypol.cnot(4.5,25.5,5.75,9)
	\mypol.cnot(4.5,25,6,9)
    \begin{pgfonlayer}{secondbackground}
	\mypol.box(-3,-0.5,10,42.5, black!5)
    \end{pgfonlayer}
\end{tikzpicture}
	\end{minipage}
	\hfill
	\begin{minipage}[b]{0.16\textwidth}
\begin{tikzpicture}[scale=0.128]
\node at (-9.4,1) {d)};
	\mystep.first(0,0)						
	\mystep.first(0,21)
	\mypol.cnot(4.5,25.5,5.75,9)
	\mypol.cnot(4.5,25,6,9)
    \begin{pgfonlayer}{secondbackground}
	\mypol.box(-7,-1.5,15,87, black!2)
    \end{pgfonlayer}
    \begin{pgfonlayer}{secondbackground}
	\mypol.box(-3,-0.5,10,42.5, black!5)
    \end{pgfonlayer}
	\mystep.first(0,43)						
	\mystep.first(0,64)
	\mypol.cnot(4.5,68.5,5.75,9)
	\mypol.cnot(4.5,68,6,9)
    \begin{pgfonlayer}{secondbackground}
	\mypol.box(-3,42.5,10,42.5, black!5)
    \end{pgfonlayer}
	\mypol.cnot(4.5,47.5,9.75,10)
	\mypol.cnot(4.5,47,10,10)
\end{tikzpicture}
	\end{minipage}
\caption[Coding for the compound channel.]{Coding for the compound channel: first four iteration steps. In a) and b) we also show the used decoding order denoted by the numbered arrows.}\label{onetwo}
\end{figure}

The main idea is to align the sets $\mathcal{A}_{\text{II}}$ and $\mathcal{A}_{\text{III}}$ such that the information can be decoded via either channel. 
For now we assume that both channels involved in the compound channel have equal capacity. Hence the sets $\mathcal{A}_{\text{II}}$ and $\mathcal{A}_{\text{III}}$ are of equal size. Therefore we combine the first channel from $\mathcal{A}_{\text{II}}$ in one polarized block with the first channel from $\mathcal{A}_{\text{III}}$ in another polarized block by an additional CNOT gate. Then we do the same with the second channel in each set and so on. After aligning all channels from these sets in such a way we have a new polarized block of size $2N$ and the relative amount of incompatible indices in this block is half of the original polarized block of size $N$. The receiver now decodes the channels in an order such that the successive cancellation decoder works. We repeat this procedure recursively and in each step the number of unaligned indices halves (see Figure \ref{onetwo}). 

Since we assume equal capacity we can align all incompatible indices and hence code at the compound capacity which is, in this case, simply equal to the capacity of the involved channels. \\

Now we look at the case when the capacities are not equal, meaning that the sets $\mathcal{A}_{\text{II}}$ and $\mathcal{A}_{\text{III}}$ are not necessarily of equal size any more.  
At some point during the alignment we will have combined all indices from the smaller set with indices from the other set, while that one has still unaligned indices. 
We handle that by simply sending frozen bits through these unaligned channels. Hence we can only code at the minimum of the capacities of the involved channels, but this is indeed the capacity of the compound channel. \\

Also we can use the same technique to code for compound channels based on sets with more than two channels. Therefore we first align two of the involved channels and get a polar code reliable for these two channels. Now we repeat the process aligning this new polarized code with the one for the third channel. This gives us a reliable code for the three involved channels. We can repeat this until the code is reliable for the ``k-member'' compound channel for which we wanted to code. 
\newpage   
\section{Polar codes for the Multiple Access Channel}\label{MAC}

\subsection{Two-user MAC} \label{2MAC}
In this section we will use polar codes to show how to code for the two-user binary-input multiple access channel at the best known rate.
Moreover we will apply techniques based on our results in \cite{HMW14}. 
The achievable rate region for the two-user classical-quantum MAC was proven by Winter in \cite{W01} and is
described by the following bounds: 
\begin{eqnarray}
R_{x} &\leq &I(X;B|Y)_\rho \\
R_{y} &\leq &I(Y;B|X)_\rho \\
R_{x}+R_{y} &\leq &I(XY;B)_\rho
\end{eqnarray}%
with respect to a ccq-state 
\begin{equation}
\rho ^{XYB}=\sum_{x,y}p_{X}(x)p_{Y}(y)\ketbra{X}{x}{x}\otimes %
\ketbra{Y}{y}{y}\otimes \rho _{x,y}^{B}.
\end{equation}

The case when the last inequality above is saturated and therefore results in a line, which interpolates between the points 
$(I(X;B)_\rho, I(Y;B|X)_\rho )$ and $(  I(X;B|Y)_\rho,I(Y;B)_\rho )$, called the \textit{%
dominant face} of the rate region, is of particular
interest to us. That is because it is clear that if every point on the
dominant face can be achieved then we can also achieve every other point
within the rate region by exploting resource wasting.

Recently, Arikan introduced the technique of ``monotone chain rules''
for handling the Slepian-Wolf problem
\cite{A12}, which is compressing two correlated classical sources without loss, 
 with the polar coding technique. Later
\"Onay applied this approach to the classical binary-input MAC \cite{O13}.
The advantage of this approach is that with each monotone chain rule,
we can achieve a rate pair lying on the dominant face of the rate region.
Furthermore, the achievable points form a dense subset of all points
on the dominant face, so that we can approximate every point on the dominant face to
arbitrarily good accuracy. 
Here we apply the technique to the classical-quantum binary-input MAC with two senders. \\

We first recall the idea of a monotone chain rule, but with our discussion here involving
the classical-quantum MAC. 
Let $X^N$ and $Y^N$ each denote a sequence of $N$ uniformly random bits. Let
$U^N$ be the result of sender 1 processing the sequence $X^N$ with the polar encoder $G_N$,
and let $V^N$ be the result of sender 2 processing the sequence $Y^N$ with the same polar encoder $G_N$.

Let $(S_1, \dots, S_{2N})$ be a permutation
of the input sequence $U^NV^N$ such that the relative order of the elements
constituting $U^N$ is preserved. A chain-rule expansion for mutual
information is said to be \emph{monotone with respect to $U^N$} if it is of the
following form:
\begin{equation}  \label{NI}
N \cdot I(XY;B) = I(U^N V^N;B^N) = \sum_{i=1}^{2N} I(S_i;B^N | S^{i-1}),
\end{equation}
with the first equality following from the reversibility of the encoders and the
second from the chain rule for mutual information.
Based on the above permutation, we let $b^{2N}$ denote a binary sequence which
we can think of as a ``path'', where $b_k$ is equal to zero
if the $k$th channel use is transmitting an information bit from the input
sequence $U^N$ of the first sender and equal to one if the $k$th channel use
is transmitting an information bit from the input sequence $V^N$ of the second sender. 
Intuitively this can be seen as following a path on a two dimensional lattice, see Figure \ref{2d}, where the monotonicity is given if our path is continuous and passes indices only in ascending order. 
\newcommand\XA{0.5}
\newcommand\Thick{3pt}
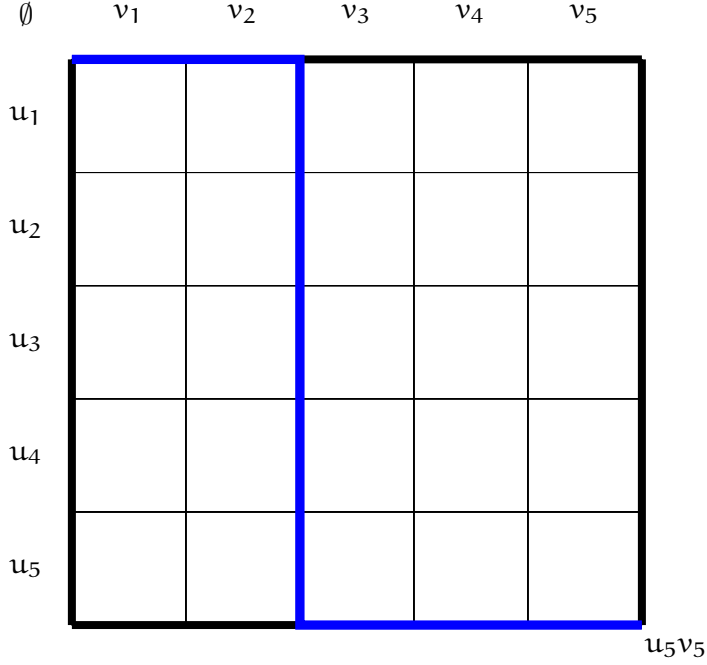
\begin{figure}
\centering
\begin{tikzpicture}[scale=1.5]
 
\foreach \z in {0,...,5}{
 \foreach \y in {0,...,5}{
	 \draw[ultra thin] (0+\y,0) -- (0+\y,5);
	 \draw[ultra thin] (0,0+\y) -- (5,0+\y);
   }
  }

   	 \draw[line width=\Thick] (0,0) -- (0,5);
	\draw[line width=\Thick] (0,0) -- (5,0);
	\draw[line width=\Thick] (5,5) -- (0,5);
	\draw[line width=\Thick] (5,5) -- (5,0);
  \foreach \x in {1,...,5}{
	\draw (-0.4, 5.5-\x)node  {$u_{\x}$};
}
  \foreach \x in {1,...,5}{
	\draw (-0.5+\x,5.4)node  {$v_{\x}$};
}
\draw (-0.4 ,5.4)node  {$\emptyset$};
\draw (5.3,-0.2)node  {$u_5v_5$};
\draw[line width=3.5pt, blue] (0,5) -- (2,5) -- (2, 0) -- (5,0);
\end{tikzpicture}
\caption[Order for decoding the 2-user MAC]{Order for decoding the 2-user MAC, with example path in blue.}\label{2d}
\end{figure}

This results in the following rates 
\begin{align}
R_x &= \frac{1}{N} \sum_{k:b_k=0} I(S_k;B^N|S^{k-1}) \leq \frac{1}{N} I(U^N ;B^N|V^N) = I(X;B|Y) \\
R_y &= \frac{1}{N} \sum_{k:b_k=1} I(S_k;B^N|S^{k-1}) \leq \frac{1}{N} I(V^N ;B^N|U^N) = I(Y;B|X) \\
R_x + R_y &= I(XY;B),
\end{align}
where each of the inequalitites hold because of the structure of the monotone chain
rules in (\ref{NI}), the statistical independence of $U^N$ and $V^N$,
and the one-to-one correspondence between $U^N, V^N$ and $X^N, Y^N$. Using the standard polar coding channel-combining technique,
outlined in the previous section, we get a combined channel $W_N$ from $W^N$ by transforming
both input sequences 
\begin{equation}
W_N(x^N,y^N) =W^N(u^N G_N,v^N G_N).
\end{equation}
In contrast to the single sender case we now have to distinguish whether we want to
decode a bit from sender 1 or from sender 2 in the split channel step. This is done as follows 
\begin{equation}
W_N^{(b_k,i,j)}= 
\begin{cases}
W_N^{(0,i,j)}: u_i \rightarrow \rho^{U^{i-1}_1 V_1^j B^N}_{(0,i,j),u_i}
\quad \quad \mathrm{if} \; b_k =0 \\ 
W_N^{(1,i,j)}: v_j \rightarrow \rho^{U^{i}_1 V_1^{j-1} B^N}_{(1,i,j),v_j}
\quad \quad \mathrm{if} \; b_k =1%
\end{cases}%
\end{equation}
with output states 
\begin{eqnarray}
\rho^{U^{i-1}_1 V_1^j B^N}_{(0,i,j),u_i} &=& \sum_{u_1^{i-1}, v_1^j} \frac{1%
}{2^{k-1}} \ketbra{}{u_1^{i-1}}{u_1^{i-1}} \otimes \ketbra{}{v_1^j}{v_1^j}
\otimes \bar\rho^{B^N}_{u_1^i,v_1^j} \\
\rho^{U^i_1 V_1^{j-1} B^N}_{(1,i,j),v_j} &=& \sum_{u_1^{i}, v_1^{j-1}} \frac{%
1}{2^{k-1}} \ketbra{}{u_1^{i}}{u_1^{i}} \otimes \ketbra{}{ v_1^{j-1}}{
v_1^{j-1}} \otimes \bar\rho^{B^N}_{u_1^i,v_1^j},
\end{eqnarray}
with 
\begin{equation}
\bar\rho^{B^N}_{u_1^i,v_1^j} = \sum_{u_{i+1}^N,v_{j+1}^N} \frac{1}{2^{2N-k}}
\rho^{B^N}_{u^N,v^N}.
\end{equation}
Similar to the case of classical-quantum polar coding for a single sender,
we now discuss how a \textit{quantum successive cancellation decoder} operates for the
cq-MAC. As in \cite{HMW14} we build a decoder based on the ideas introduced in \cite{WG13} using projectors to decide whether the $k$%
th input, corresponding to the split channel $W_N^{(b_k,i,j)}$, is equal to zero or
one: 
\begin{equation}
\Pi_{(b_k,i,j),0}= 
\begin{cases}
\Pi^{U^{i-1}_1 V_1^j B^N}_{(0,i,j),0} \quad \quad \mathrm{if} \; b_k =0 \\ 
\Pi^{U^i_1 V_1^{j-1} B^N}_{(1,i,j),0} \quad \quad \mathrm{if} \; b_k =1%
\end{cases}%
\end{equation}
with 
\begin{align}
\Pi^{U^{i-1}_1 V_1^j B^N}_{(0,i,j),0} = \left\{ \sqrt{\rho^{U^{i-1}_1 V_1^j
B^N}_{(0,i,j),0}} - \sqrt{\rho^{U^{i-1}_1 V_1^j B^N}_{(0,i,j),1}} \geq 0 \right\}
\\
\Pi^{U^i_1 V_1^{j-1} B^N}_{(1,i,j),0} = \left\{ \sqrt{\rho^{U^i_1 V_1^{j-1}
B^N}_{(1,i,j),0}} - \sqrt{\rho^{U^i_1 V_1^{j-1} B^N}_{(1,i,j),1}} \geq 0 \right\}
\end{align}
and 
\begin{equation}
\Pi_{(b_k,i,j),1} = \id - \Pi_{(b_k,i,j),0}.
\end{equation}
With ${A\geq 0}$ we denote the projector onto the positive eigenspace of $A$ and
with ${A < 0}$ the projector onto its negative eigenspace.

Note that we cannot directly build a POVM from these measurements since their dimension depends on $i$. Thus, similar to \cite{WG13}, we write 
\begin{align}
\Pi^{U^{i-1}_1 V_1^j B^N}_{(0,i,j),0} =& \sum_{u_1^{i-1}, v_1^j} \ketbra{}{u_1^{i-1}}{u_1^{i-1}} \otimes \ketbra{}{v_1^j}{v_1^j}\otimes \Pi^{B^N}_{(0,i,j),u_1^{i-1} 0,v_1^j} \\
\Pi^{U^i_1 V_1^{j-1} B^N}_{(1,i,j),0} =& \sum_{u_1^{i}, v_1^{j-1}} \ketbra{}{u_1^{i}}{u_1^{i}} \otimes \ketbra{}{ v_1^{j-1}}{
v_1^{j-1}} \otimes \Pi^{B^N}_{(1,i,j),u_1^i, v_1^{j-1}0}
\end{align}
where
\begin{align}
\Pi^{B^N}_{(0,i,j),u_1^{i-1}0, v_1^j } =& \left\{ \sqrt{\bar\rho^{B^N}_{u_1^{i-1}0,v_1^j}} - \sqrt{\bar\rho^{B^N}_{u_1^{i-1}1,v_1^j}} \geq 0 \right\}
\\
\Pi^{B^N}_{(1,i,j),u_1^i, v_1^{j-1}0} =& \left\{ \sqrt{\bar\rho^{B^N}_{u_1^i,v_1^{j-1}0}} - \sqrt{\bar\rho^{B^N}_{u_1^i,v_1^{j-1}1}} \geq 0 \right\}.
\end{align}

These measurements can be used to construct a POVM, with arguments presented in \cite{WG13}, with
elements 
\begin{align}
\Lambda_{u^N,v^N} = \Pi_{(b_1,i_1,j_1),\{u_1, v_1\} } &\cdots \Pi_{(b_{2N},i_{2N},j_{2N}),\{u_1^{N-1}u_Nv_1^N, u_1^Nv_1^{N-1}v_N\} } \cdots \nonumber\\
&\cdots \Pi_{(b_1,i_1,j_1),\{u_1,
v_1\} } ,
\end{align}
where the exact value of $i$ and $j$ depend on the monotone chain rule
chosen for decoding, as well as whether a projector attempts to
decode $u_i$ or $v_j$. As required for a POVM we also have
\begin{equation}
\sum_{u_A, v_A} \Lambda_{u^N,v^N} = \id^{B^N},
\end{equation}
by noting that we can set $\Pi_{(b_k,i,j),\{u_i, v_j\} } = \id$ when $\{u_i,
v_j\}$ is a frozen bit.

Using the bitwise projections we can build the successive cancellation
decoder with the decoding rules: 
\begin{align}
\hat u_i =&
\begin{cases}
u_i \quad \quad & \mathrm{if} \; i\in A^c \\ 
h(\hat u_1^{i-1}, \hat v_1^j) \quad \quad & \mathrm{if} \; i\in A%
\end{cases}
\\
\hat v_j =& 
\begin{cases}
v_j \quad \quad & \mathrm{if} \; j\in A^c \\ 
g(\hat u_1^{i}, \hat v_1^{j-1}) \quad \quad & \mathrm{if} \; j\in A,
\end{cases}%
\end{align}
where $h(\hat u_1^{i-1}, \hat v_1^j)$ is the outcome of the $k$th
measurement when $b_k=0$ based on 
\begin{equation}
\{ \Pi^{B^N}_{(0,i,j),u_1^{i-1}0, v_1^j } , \Pi^{B^N}_{(0,i,j),u_1^{i-1}1, v_1^j } \}
\end{equation}
and $g(\hat u_1^{i}, \hat v_1^{j-1})$ is the outcome when $b_k=1$ based on 
\begin{equation}
\{ \Pi^{B^N}_{(1,i,j),u_1^i, v_1^{j-1}0} , \Pi^{B^N}_{(1,i,j),u_1^i, v_1^{j-1}1} \}.
\end{equation}

Due to the structure of the decoder and the polarization effect,
the block error probability
decays exponentially with $o(2^{-\frac{1}{2}N^\beta})$ where $\beta< \frac{1}{2}$ and $N$ the number of channel uses as in the single-sender case described in the previous section.

\paragraph{Continuity of rates and approximations}\label{rates}

In this paragraph we will argue that the above approach can be used to achieve
the entire dominant face of the rate region. This task was achieved for the
classical Slepian-Wolf source coding problem \cite{A12} involving rates
based on the conditional Shannon entropy, and then extended to the complementary problem
of channel coding over the classical MAC \cite{O13}. We present the extension of this technique from \cite{HMW14}, where we applied it to channel coding for the classical-quantum MAC.

First we define a distance measure for comparing paths. Let $b^{2N},\tilde b^{2N}$ denote
two paths and $(R_u, R_v), (\tilde R_u, \tilde R_v)$ their corresponding
rate pairs. Then we define the distance between the two paths $b^{2N}$ and $\tilde b^{2N}$ as follows 
\begin{equation}
d(b^{2N},\tilde b^{2N}) \equiv |R_u - \tilde R_u | = |R_v - \tilde R_v |
\end{equation}
where the last equality holds since $R_u +R_v = \tilde R_u +\tilde R_v =
I(XY;B)$.

We now define two paths $b^{2N},\tilde b^{2N}$ to be neighbors if $\tilde
b^{2N}$ can be obtained from $b^{2N}$ by transposing $b_i$ with $b_j$ for
some $i<j$ such that $b_i\neq b_j$ and $b_{i-1}^{j+1}$ is either all 0 or
all 1. We state the following proposition from \cite{HMW14} bounding the distance between two
neighboring paths. This generalizes Proposition 3 in \cite{A12} to the
quantum setting considered here. 
\begin{prop}\label{neig}
If paths $b^{2N}$ and $\tilde b^{2N}$ are neighboring, then the following holds 
\begin{equation}
d(b^{2N},\tilde b^{2N}) \leq \frac{1}{N}.
\end{equation}
\end{prop}

We prove the above proposition in \cite{HMW14}, where we generalize the proof for the classical setting in \cite{A12}. 

The main idea in the proof is to upper bound the difference in the rates corresponding to the two neighboring paths. 

This can easily be done by checking the corresponding chain rules for the four possible cases of substrings $( b_i^j \in \{ 01^{j-i}, 0^{j-i}1, 1^{j-i}0, 10^{j-i} \} )$, dependent on the coordinates $i<j$, transposition of which defines the difference of the neighboring paths. \\

Using this result we can show that the distribution of achievable points on
the dominant face of the rate region is dense, and we state the following
theorem, again from \cite{HMW14}, generalizing Theorem 1 in \cite{A12} to the quantum case considered here. 
\begin{thm}\label{thm}
Let $(R_x, R_y)$ be a given rate pair on the dominant face. For any given $\epsilon >0$, there exists an $N \in \mathbb{N}$ and a chain rule $b^{2N}$ on $U^NV^N$ such that $b^{2N}$ belongs to the class $\nu_{2N}  = \{ 0^i 1^N 0^{N-i} : 0 \leq i \leq N \}$ and has a rate pair $(R_1, R_2)$ satisfying 
\begin{equation}
|R_1 - R_x| \leq \epsilon \qquad\textit{and}\qquad |R_2 - R_y| \leq \epsilon.
\end{equation}
\end{thm}
As shown in \cite{HMW14} the proof follows from the fact that in $\nu_{2N}$ the two paths $0^i 1^N 0^{N-i}
$ and $0^{i+1} 1^N 0^{N-i-1}$ are neighbors, thus we simply have to fix $N > 
\frac{1}{\epsilon}$ and Theorem \ref{thm} follows from Proposition \ref{neig}.

Note that this result is only concerned with permutations from the class $%
\nu_{2N}$. This is sufficient for our purposes, but it may be
interesting to consider a more general class when choosing the decoding path.

\paragraph{Path scaling and polarization}\label{scaling}

It was shown in the previous section that one can find polar codes that
approximate points on the dominant face of the rate region for any cq-MAC.
Here we show that performing a step of the polar coding
recursion does not change the achievable rates so long as polarization still
holds, because these approximations are stable under scaling of the
chosen path. These are generalizations from \cite{HMW14} of ideas from \cite{A12} to the classical-quantum MAC.

We look at paths $kb^{2N}$ which denote the scaling of a path $b^{2N}$ as 
\begin{equation}
\underbrace{b_{1}\dots b_{1}}_{k}\underbrace{b_{2}\dots b_{2}}_{k}\dots
\dots \underbrace{b_{2N}\dots b_{2N}}_{k}
\end{equation}%
and therefore represent a monotone chain rule for $U^{kN}V^{kN}$. Note that
we can write a step of the polar-code transformation as 
\begin{equation}
T_{2i-1}=S_{i}\oplus \tilde{S}_{i},\;T_{2i}=S_{i},
\end{equation}%
and thus we can show that an additional step of polarization does not affect
the rate 
\begin{align}
\begin{aligned}
I(T_{2i-1};B^{N}|T^{2i-2}) +& I(T_{2i};B^{N}|T^{2i-1}) \\
&=I(T_{2i-1}T_{2i};B^{N}|T^{2i-2}) \\
&=I(S_{i}\oplus \tilde{S}_{i},S_{i};B^{N}|S^{i-1}\oplus \tilde{S}%
^{i-1},S^{i-1}) \\
&=I(\tilde{S}_{i},S_{i};B^{N}|\tilde{S}^{i-1},S^{i-1}) \\
&=2I(S_{i};B^{N}|S^{i-1})
\end{aligned}
\end{align}
where the last step follows if $T^{4N}$ follows the path $2b^{2N}$. From
this we can conclude that if a path $b^{2N}$ achieves a rate $(R_{1},R_{2})$
then the path $2b^{2N}$ achieves the same rate pair.

With the above observations, the polarization argument follows directly using arguments from the
single-sender case in Section \ref{cq}.

\paragraph{Polar code performance}

Since we use a POVM with the same basic structure as the single-sender case,
the analysis of the error probability follows the same arguments.

That is, applying Gao's union bound \cite{J14} and our results in Section \ref{imp}
to the probability of error $P_e(N, b^{2N}, (K_u,K_v), (A_u, A_v), (u_{A_u^c}, v_{A_v^c}))$ for code length $N$, a chosen path $b^{2N}$, the number of information bits $(K_u,K_v)$ and sets of information bits $(A_u, A_v)$ for each sender and the choice for the frozen bits $(u_{A_u^c}, v_{A_v^c})$,
we get 
\begin{equation}
P_e(N, b^{2N}, (K_u,K_v), (A_u, A_v), (u_{A_u^c}, v_{A_v^c})) 
\leq 4 \sum_{i\in A_u, j\in A_v} \frac{1}{2} \sqrt{ F(W_N^{(b_k,i,j)})}
\end{equation}
and therefore we can state that the error probability in Section \ref{imp} also
holds for multiple-user settings 
\begin{equation}
P_e(N, R) = o(2^{-N^\beta}).
\end{equation}
This result is an improvement over our previous result in \cite{HMW14}, in analogy to the improvement in Section \ref{imp}. 

\subsection{k-user MAC}

\label{3MAC} 
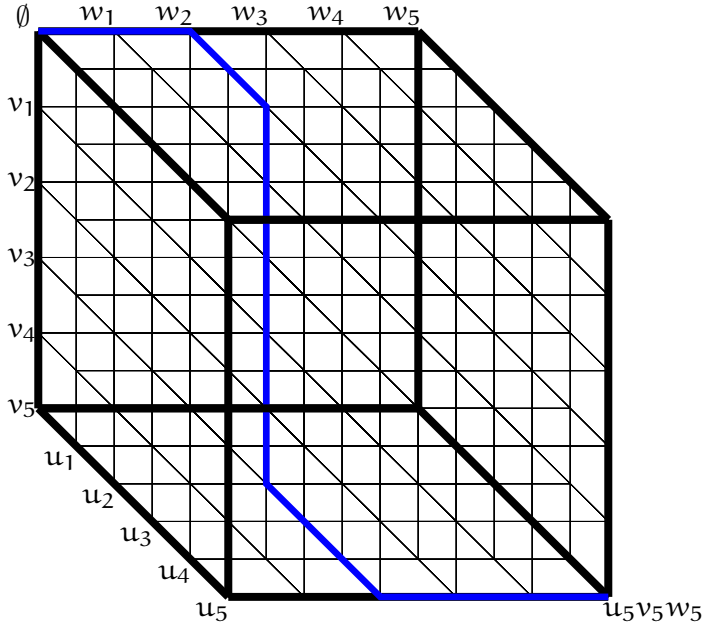
\begin{figure}[ht]
\centering
\begin{tikzpicture}[scale=1]
\foreach \z in {0,...,5}{
 \foreach \y in {0,...,5}{
  \foreach \x in {0,...,5}{
   	 \draw[ultra thin] (\x-\XA*\y,0+\XA*\y) -- (\x-\XA*\y,5+\XA*\y);
 	 \draw[ultra thin] (0-\XA*\y,\x+\XA*\y) -- (5-\XA*\y,\x+\XA*\y);
	 \draw[ultra thin] (0+\y,0+\z) -- (-5*\XA+\y,5*\XA+\z);
   }
  }
}
   	 \draw[line width=\Thick] (0,0) -- (0,5);
	\draw[line width=\Thick] (0,0) -- (5,0);
	\draw[line width=\Thick] (5,5) -- (0,5);
	\draw[line width=\Thick] (5,5) -- (5,0);
	\draw[line width=\Thick] (5,5) -- (2.5,7.5);
	\draw[line width=\Thick] (-2.5,2.5) -- (-2.5,7.5);
	\draw[line width=\Thick] (2.5,7.5) -- (-2.5,7.5);
	\draw[line width=\Thick] (2.5,7.5) -- (2.5,2.5);
	\draw[line width=\Thick] (0,5) -- (-2.5,7.5);
	\draw[line width=\Thick] (-2.5,2.5) -- (0,0);
	\draw[line width=\Thick] (5,0) -- (2.5,2.5);
	\draw[line width=\Thick] (2.5,2.5) -- (-2.5,2.5);
  \foreach \x in {1,...,5}{
	\draw (-2.7 + \x*0.5,2.3 - \x*0.5)node  {$u_{\x}$};
}
  \foreach \x in {1,...,5}{
	\draw (-2.7,7.5 - \x)node  {$v_{\x}$};
}
  \foreach \x in {1,...,5}{
	\draw (-2.7 + \x,7.7)node  {$w_{\x}$};
}
\draw (-2.7 ,7.7)node  {$\emptyset$};
\draw (5.6,-0.2)node  {$u_5v_5w_5$};
\draw[line width=2.5pt, blue] (-2.5, 7.5) -- (-0.5, 7.5) -- (0.5, 6.5) -- (0.5, 1.5) -- (2, 0) -- (5, 0);
	\draw[line width=\Thick] (5,5) -- (0,5);
	\draw[line width=\Thick] (2.5,2.5) -- (-2.5,2.5);
\end{tikzpicture}
\caption[Order for decoding the 3-user MAC.]{Order for decoding the 3-user MAC, with example path in blue. Allowed paths are those which go only to the right, front or bottom at each step, to imply a monotone chain rule. This is important to ensure the correct decoding order for the successive cancelation decoder.}\label{3dpath}
\end{figure}

The approach for the two sender MAC discussed above can easily be extended to the
case of many senders. 
We will first review the case of three senders as we have described in \cite{HMW14}, because we will especially need these techniques for the classical-quantum interference channel in the next Section. 
To generalize for three senders we follow an approach similar to the two-dimensional case, by following a path through a
three-dimensional cube, see Figure \ref{3dpath} for example,
in order to choose a path with $b_k\in \{0,1,2\}$ giving rise to the following
achievable rate region 
\begin{align}
R_x = \frac{1}{N} \sum_{k:b_k=0} I(S_k;B^N|S^{k-1}) \leq I(X;B|YZ) \\
R_y = \frac{1}{N} \sum_{k:b_k=1} I(S_k;B^N|S^{k-1}) \leq I(Y;B|XZ) \\
R_z = \frac{1}{N} \sum_{k:b_k=2} I(S_k;B^N|S^{k-1}) \leq I(Z;B|XY) \\
R_x + R_y \leq I(XY;B|Z) \\
R_x + R_z \leq I(XZ;B|Y) \\
R_y + R_z \leq I(ZY;B|X) \\
R_x + R_y + R_z = I(XYZ;B).
\end{align}

Note that the achievable rate region and the reliability of the code follow from the same arguments as in the two-sender case in Section \ref{2MAC}. \\

To generalize the method to the k-user MAC we simply have to follow a path with $b_k\in \{0,\dots,k-1\}$ such that the entropy inequalities defining the achievable rate region remain monotonic chain rules. 
This, in analogy to the previous approaches, can be seen as following a continuous monotonic path in a $k$-dimensional structure. 

Again the achievable rate region and reliability follow from arguments similar to the two-sender case.
\newpage   
\section{Universal polar codes for the compound MAC}\label{comMAC}

\label{cMAC}

In this section we will describe how the universal polar codes from Section \ref{cqUChannel} and the coding for the classical-quantum multiple access channel in Section \ref{MAC} can be combined to get reliable polar codes for the compound MAC. 
We will therefore follow our results in \cite{HMW14} generalizing the second scheme
described in \cite{HU13} and the generalizations of this scheme to MACs in 
\cite{WS14}. 

For now we will look at compound MACs based on sets of two different MACs and, as in the single sender case, we assume that the chosen channel is known to the receiver but not to the sender. 
We also start by considering a compound MAC described by a set of two MACs with equal sum rate. 
Both assumptions, that is, the limit on the number of channels and on the sum rate, will be lifted later. 
The essential approach, similar to that of single sender channels, is to \textquotedblleft align\textquotedblright\
polarized indices.
The approach in Section \ref{MAC} will allow us to align both senders independently. 

Recall that in Section \ref{cq} we reviewed the channel splitting step for classical-quantum channel polarization, first introduced in \cite{WG13}. Similarly, here we define the partial split channels 
\begin{align} 
\begin{aligned}
P_{i} &: U_{i}\rightarrow B^{N}U_{1}^{i-1}V_{1}^{j} \\
Q_{i} &: U_{i}\rightarrow B^{N}U_{1}^{i-1}V_{1}^{j},
\end{aligned}
\end{align}
each corresponding to the first sender of one of the two MACs in the set comprising the compound MAC. In the previous section, $b_k$ served as a label indicating which sender should be decoded in the $k$th step.
Moreover, the channels $P_i$ and $Q_i$ can be considered to be equivalent to looking at only the channel uses of the corresponding MACs for which $b_k=0$. 
Similar to Equation \ref{GGBB} we define
\begin{equation}\label{sets}
\begin{split}
\mathcal{G}_{(1)}& =\left\{i\in \lbrack 1:N]:\sqrt{F(P_{i})}<2^{-N^{\beta }}\right\}, \\
\mathcal{G}_{(2)}& =\left\{i\in \lbrack 1:N]:\sqrt{F(Q_{i})}<2^{-N^{\beta }}\right\}, \\
\mathcal{B}_{(1)}& =\left\{i\in \lbrack 1:N]:\sqrt{F(P_{i})}\geq 2^{-N^{\beta }}\right\}, \\
\mathcal{B}_{(2)}& =\left\{i\in \lbrack 1:N]:\sqrt{F(Q_{i})}\geq 2^{-N^{\beta }}\right\}.
\end{split}%
\end{equation}%
denoting the sets of indices corresponding to whether a bit is good or bad for a
channel. These sets tell us whether the attempt to send an information
bit through one of the MACs would be successful with high probability for
the $i$th channel use of the first sender. Due to the polarization
effect, as in Equation \ref{AAAA}, all bits will be in one of the following sets: 
\begin{align}\label{csets}
\mathcal{A}_{\text{I}}\ & =\mathcal{G}_{(1)}\cap \mathcal{G}_{(2)}, \\
\mathcal{A}_{\text{II}}\ & =\mathcal{G}_{(1)}\cap \mathcal{B}_{(2)}, \\
\mathcal{A}_{\text{III}}& =\mathcal{B}_{(1)}\cap \mathcal{G}_{(2)}, \\
\mathcal{A}_{\text{IV}}& =\mathcal{B}_{(1)}\cap \mathcal{B}_{(2)}.
\end{align}%
Bits belonging to the sets $\mathcal{A}_{\text{I}}$ will also be decoded
with high probability in the compound setting and bits in $\mathcal{A}_{%
\text{IV}}$ will have to be set as frozen bits. 

Due to Theorem \ref{thm} in Section \ref{rates} we can find monotone chain rules for each MAC which approximate every point on the dominant face of the rate region of the corresponding MAC.

The main idea from here is to align the sets $\mathcal{A}_\text{II}$ and $\mathcal{A}_%
\text{III}$ within a recursion to achieve the capacity of the compound MAC. We will do so alternating in each step of
the recursion either for the first or the second sender. Here, just as in
the classical case, we have to ensure that we align bits such that the
successive cancellation decoder can still be applied.

We take two polar coding blocks which have both already been
polarized independently of each other. Since both blocks have been built
from the same channel, the sets of indices are identical for both blocks. We
then combine the first index from $\mathcal{A}_\text{II}$ in the first block
with the first index of $\mathcal{A}_\text{III}$ in the second block by an
additional CNOT gate, and similarly for the second indices and so on. With such a scheme,
we can halve the fraction of incompatible indices, those from
the sets $\mathcal{A}_\text{II}$ and $\mathcal{A}_\text{III}$, for the first
sender.

Intuitively this can be seen as sending the same information bits via both
of the aligned channels, so that the receiver will be able to decode one of
them independently of which MAC is actually used. Since we assume that the
receiver knows the applied channel, this works well with the successive
cancellation decoder, because the receiver can just decode the channel which
is good for the used MAC first and then decode the aligned channel as if it
is a frozen bit.

In the next step we take two of the blocks after the first iteration step
and repeat the process for the second sender. Hence, we again halve the
fraction of incompatible indices for this sender. In the following we
repeat this process until the fraction of incompatible indices tends to
zero.

\pgfooclass{stamp}{     \method stamp() {     }
 \method cnot(#1,#2,#3,#4) {         	\draw (#1,#2) -- (#1-#3,#2) -- (#1-#3,#2-#4) -- (#1,#2-#4);
	\draw (#1-#3,#2) circle (0.1);
	\draw (#1-#3,#2+0.1) -- (#1-#3,#2) -- (#1-#3-0.25,#2);
	\draw[dotted] (#1-#3-0.25,#2) -- (#1-#3-0.75,#2);
   }
} \pgfoonew \mystamp=new stamp()

\begin{figure}[tb]
\centering
\begin{tikzpicture}[scale=1]
	\draw (0,0) -- (1,0) -- (1,1) -- (0,1) -- cycle; 
	\node[font=\tiny] at (0.5,0.5) {$U_{2b}$};
	\draw (0,1) -- (1,1) -- (1,3) -- (0,3) -- cycle; 
	\node[font=\tiny] at (0.5,2) {$V_2$};
	\draw (0,3) -- (1,3) -- (1,4) -- (0,4) -- cycle; 
	\node[font=\tiny] at (0.5,3.5) {$U_{2a}$};
	\draw (0,5) -- (1,5) -- (1,6) -- (0,6) -- cycle; 
	\node[font=\tiny] at (0.5,5.5) {$U_{1b}$};
	\draw (0,6) -- (1,6) -- (1,8) -- (0,8) -- cycle; 
	\node[font=\tiny] at (0.5,7) {$V_1$};
	\draw (0,8) -- (1,8) -- (1,9) -- (0,9) -- cycle; 
	\node[font=\tiny] at (0.5,8.5) {$U_{1a}$};
	\mystamp.cnot(0,8.7,1,8);
	\mystamp.cnot(0,8.4,1.3,8);
           \draw[->,solid] (-0.5,9) node[font=\tiny, below right=-0.1]{1} -- (-0.5,8.7); 			
	\draw[->,densely dotted] (-0.5,8.7)node[font=\tiny, below right=-0.1]{2} -- (-0.5,8.4);
	\draw[->] (-0.5,8.4)node[font=\tiny, below right=-0.1]{3} -- (-0.5,5);
          \draw[->] (-0.5,4)node[font=\tiny, below right=-0.1]{1} -- (-0.5,0.7);			 
	\draw[->,densely dotted] (-0.5,0.7)node[font=\tiny, below right=-0.1]{2} -- (-0.5,0.4);
	\draw[->] (-0.5,0.4)node[font=\tiny, below right=-0.1]{3} -- (-0.5,0);
\end{tikzpicture}
\caption[Coding for the compound MAC.]{Coding for the compound MAC: first iteration step. The arrows indicate
the decoding order and must be followed in increasing order of the attached
numbers.}
\label{decMAC}
\end{figure}
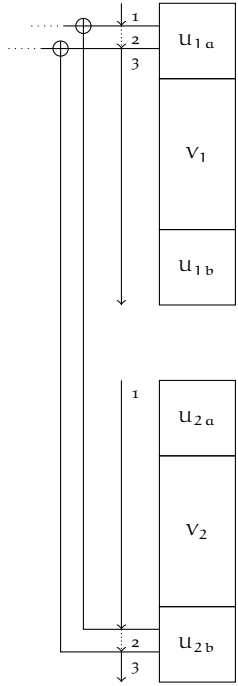

To generalize the above scheme to the $k$-user MAC we simply decode by
alternating over the different senders in each step of the recursion.
Therefore each sender becomes aligned in every $k$-th step of the recursion.
In the recursion $mk$ steps reduce the fraction of incompatible indices for
the $k$-user MAC to $(|\mathcal{A}_\text{II}| + |\mathcal{A}_\text{III}|)/2^mN$%
. For example, for the case of two users this means that every two steps we
halve the fraction of incompatible indices. This is due to the fact that we
need one recursion step for each sender to halve the fraction of incompatible
indices for that particular sender. Then we can reorder our decoding in a
way that the successive cancellation decoder still holds. Figure \ref{decMAC}
illustrates the process.

In order to use the compound MAC for the interference channel we need to generalize the described approach to the setting which includes unequal sum rates. 
Therefore assume that we want to code for a rate pair $(R_x,R_y)$ on the dominant face of the achievable rate region for the compound MAC consisting of a set of two MACs. 
Now we can find a rate pair  $(R'_x,R'_y)$ for the first MAC in that set and a rate pair $(R''_x,R''_y)$ for the second MAC such that 
\begin{align}
R_x &\leq \min(R'_x,R''_x) \\
R_y &\leq \min(R'_y,R''_y).
\end{align}
We can use the corresponding monotone chain rules for these two MACs to code for the targeted point on the rate region of the compound MAC.
Note that in the setting of unequal sum-rate the sets $\mathcal{A}_\text{II}$ and $\mathcal{A}_\text{III}$ are not necessarily of equal size.
This is not a problem for the aligning process, because we can simply align until one of the sets has no unaligned indices left and then handle the remaining indices in the larger set as frozen indices. It is easy to see that this is sufficient to code at rates $\min(R'_x,R''_x)$ and $\min(R'_y,R''_y)$ and therefore achieve the dominant face for the achievable rate region of the compound MAC.\\

Raising the limit on the size of the set of MACs constituting the compound MAC works as in the single sender case in Section \ref{cqUChannel}. 
We simply align two of the MACs from the set to get a reliable code for those and then align this new code with the third MAC. 
We can iterate this procedure to get a reliable code for a compound MAC consisting of a set of $k$ MACs. 

It was previously unknown how to code for the $k$-user compound MAC in a
quantum setting. Having this result will allow us to also code for
classical-quantum interference channels, as discussed in the next section.
\newpage   
\section{Interference channel}

\label{int}

The two-user classical-quantum interference channel \cite{SFWSH11, FHSSW12}, as illustrated in Figure \ref{fig:square}, 
and discussed in Section \ref{cqMulti}, can be fully represented by
its set of output states as follows:
\begin{equation}
\{\rho _{x_{1},x_{2}}^{B_{1}B_{2}}\}_{x_{1}\in \mathcal{X}_{1},x_{2}\in 
\mathcal{X}_{2}}.
\end{equation}

In the classical setting the best known achievable rate region for the
interference channel is given by the Han-Kobayashi rate region \cite{HK81}. We now show how we achieved this region for the two-user
classical-quantum interference channel in \cite{HMW14} using polar codes. The scheme is a
direct generalization of that presented in \cite{WS14}. Indeed the
Han-Kobayashi rate region can be achieved by splitting the message of the
first sender $m_{1}$ into two parts labelled $(l_{1},l_{2})$ and similarly
for the second sender the message $m_{2}$ is split into $(l_{3},l_{4})$. Now
we get channel inputs represented by the random variables $X_{1}^{N}$ and $%
X_{2}^{N}$ via symbol-to-symbol encoding maps $x_{1}(v_{1},v_{2})$ and $%
x_{2}(v_{3},v_{4})$ corresponding to the codewords $v_{j}^{N}(L_{j})$. Then
receiver 1 decodes from $\rho _{x_{1},x_{2}}^{B_{1}}$ the messages $%
(l_{1},l_{2},l_{3})$ and receiver 2 decodes $\rho _{x_{1},x_{2}}^{B_{2}}$ to
get the message triple $(l_{2},l_{3},l_{4})$. The Han-Kobayashi rate region
is defined for rates $(S_{1},S_{2},T_{1},T_{2})$ as follows:
\begin{align}
\begin{aligned}
S_{1}& \leq I(V_{1};B_{1}|V_{3}V_{4}), \\
T_{1}& \leq I(V_{3};B_{1}|V_{1}V_{4}), \\
T_{2}& \leq I(V_{4};B_{1}|V_{1}V_{3}), \\
S_{1}+T_{1}& \leq I(V_{1}V_{3};B_{1}|V_{4}), \\
S_{1}+T_{2}& \leq I(V_{1}V_{4};B_{1}|V_{3}), \\
T_{1}+T_{2}& \leq I(V_{3}V_{4};B_{1}|V_{1}), \\
S_{1}+T_{1}+T_{2}& \leq I(V_{1}V_{3}V_{4};B_{1}), \\
S_{2}& \leq I(V_{2};B_{2}|V_{3}V_{4}), \\
T_{1}& \leq I(V_{3};B_{2}|V_{2}V_{4}), \\
T_{2}& \leq I(V_{4};B_{2}|V_{2}V_{4}), \\
S_{2}+T_{1}& \leq I(V_{2}V_{3};B_{2}|V_{4}), \\
S_{2}+T_{2}& \leq I(V_{2}V_{4};B_{2}|V_{3}), \\
T_{1}+T_{2}& \leq I(V_{3}V_{4};B_{2}|V_{2}), \\
S_{2}+T_{1}+T_{2}& \leq I(V_{2}V_{3}V_{4};B_{2}).
\end{aligned}
\end{align}
The achievable rate region for the interference channel is the set of all
rates $(S_{1}+T_{1},S_{2}+T_{2})$.

\begin{figure}[t]
\centering
\begin{tikzpicture}[scale=0.75,baseline=(current bounding box.north)]
	\draw (0,0) -- (1,0) -- (1,1) -- (0,1) -- cycle;
	\draw (0,2) -- (1,2) -- (1,3) -- (0,3) -- cycle;
	\draw (3,0) -- (3,3) -- (2,3) -- (2,0) -- cycle;
	\draw (5,0) -- (4,0) -- (4,1) -- (5,1) -- cycle;
	\draw (5,2) -- (4,2) -- (4,3) -- (5,3) -- cycle;
          \node[font=\tiny] at (0.5,0.5) {$E_2$};
         \node[font=\tiny] at (0.5,2.5) {$E_1$};
         \node[font=\tiny] at (2.5,1.5) {$\NN_{IC}$};
         \node[font=\tiny] at (4.5,0.5) {$D_2$};
         \node[font=\tiny] at (4.5,2.5) {$D_1$};
	\draw (-1,0.5) -- (0,0.5);
	\draw (-1,2.5) -- (0,2.5);
	\draw (1,0.5) -- (2,0.5);
	\draw (1,2.5) -- (2,2.5);
	\draw (3,0.5) -- (4,0.5);
	\draw (3,2.5) -- (4,2.5);
	\draw (5,0.5) -- (6,0.5);
	\draw (5,2.5) -- (6,2.5);
         \node[font=\tiny] at (-1.5,0.5) {$M_2$};
         \node[font=\tiny] at (-1.5,2.5) {$M_1$};
         \node[font=\tiny] at (6.5,0.5) {$\hat M_2$};
         \node[font=\tiny] at (6.5,2.5) {$\hat M_1$};
         \node[font=\tiny] at (1.5,0.7) {$X_2$};
         \node[font=\tiny] at (1.5,2.7) {$X_1$};
         \node[font=\tiny] at (3.5,0.7) {$B_2$};
         \node[font=\tiny] at (3.5,2.7) {$B_1$};
\end{tikzpicture}
\caption[2-user interference channel.]{2-user interference channel. $M_i$ denotes the message sent by the $i$th sender, using an encoding map $E_i$. The $i$th receiver attempts to reconstruct the message using the decoding map $D_i$.}
    \label{fig:square}
\end{figure}
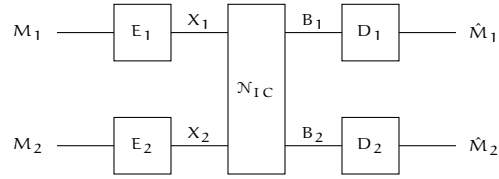
\begin{figure}[t]
\centering
\begin{tikzpicture}[scale=1.2,baseline=(current bounding box.north)]
\draw[->, >=latex] (0,0) -- (4,0) node[below]{$R_1$};
\draw[->, >=latex] (0,0) -- (0,4) node[left]{$R_2$}; 
	\draw (0,3) -- (1,3) -- (2.5,1.5) -- (2.5,0);
	\draw (0,2.5) -- (2,2.5) -- (3.5,1) -- (3.5,0);
	\filldraw[fill=blue!7] (0,2.5) -- (1.5,2.5) -- (2.5,1.5) -- (2.5,0) -- (0,0) -- cycle;
         \node[font=\tiny] at (1,3.2) {MAC 1};
         \node[font=\tiny] at (3.9,1) {MAC 2};
         \node[font=\tiny] at (1,1) {IC};
\end{tikzpicture}
\caption[Rate region of the interference channel.]{Rate region of the interference channel, induced by two MAC regions.}
    \label{fig:rect}
\end{figure}
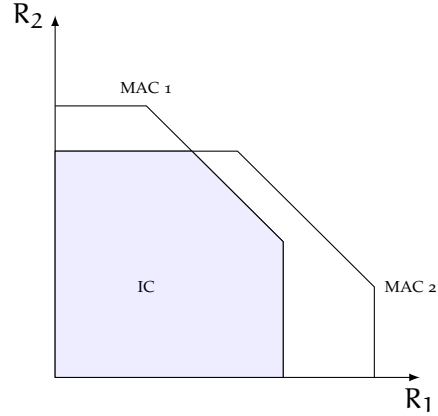

This can be seen as an interference network with four senders $x_1(v_1, v_2)$
and $x_2(v_3, v_4)$ and two recievers. The rate tuples $(S_1, T_1, T_2)$ and 
$(S_2, T_1, T_2)$ coincide with two 3-user MAC regions with two common
senders and the intersection of both gives the rate region $(S_1, S_2, T_1,
T_2)$, see Figure \ref{fig:rect}. We can find monotone chain rules for every point on the dominant
face of each MAC. Now we use the approach for the compound MAC to align the
two common senders along with the results of the previous section in
order to achieve the rate region for the two-user interference channel using
polar coding. Note that it is necessary to use the approach presented for
compound MACs in order to ensure that the decoding used for the polar codes
is good for both MACs. With this approach we can achieve the Han-Kobayashi
rate region by a successive cancelation decoder for each receiver.
There are known techniques for handling arbitrary input distributions. Rather than
go into detail, we point to the work in \cite{WS14} which elaborates on this point.
\newpage   
\section{Broadcast channel} \label{cqBroadcast}

In classical information theory there exist several schemes to show achievability of certain rate regions for the broadcast channel. 
Two of these are superposition coding \cite{B73, C72} and binning \cite{M79}. 
For the binning scheme, independent messages are sent simultaneously to both receivers and can be decoded by the respective receiver according to an argument based on joint typicality. For the superposition scheme we exploit the fact that some inputs are decodable by both receivers, see more detailed description below. 
Both techniques were combined by Marton \cite{M79} to send private messages over a broadcast channel to two receivers and achieve the so-called Martons region. 

Later this result was extended by Gelfand and Pinsker \cite{GP80} to the setting where a common message can be sent to both receivers, resulting in the so-called Marton-Gelfand-Pinsker region with common messages. 
Interestingly, it can be shown in the classical setting that even when we set the common rate in the Marton-Gelfand-Pinsker region to zero the resulting rate region is, in some cases, larger then the Martons region \cite{GK11}. \\

Broadcast channels have also been generalized to the setting of classical-quantum communication \cite{YHD11, SW12, RSW14}. 
To date the best known rate region for the classical-quantum channel has been described by Savov and Wilde in \cite{SW12} and it is a generalization of the Martons region in the classical case, and given by
\begin{align}
\begin{split}
R_1 &\leq I(U_1; B_1)_\sigma ,\\
R_2 &\leq I(U_2; B_2)_\sigma ,\\
R_1 + R_2 &\leq I(U_1; B_1)_\sigma + I(U_2;B_2)_\sigma - I(U_1;U_2)_\sigma
\end{split}
\end{align}
with respect to the state 
\begin{equation}
\sigma^{U_1U_2B_1B_2} = \sum_{u_1,u_2} p(u_1,u_2) \ketbra{}{u_1}{u_1}^{U_1} \otimes \ketbra{}{u_2}{u_2}^{U_2} \otimes \rho_{f(u_1,u_2)}^{B_1B_2},
\end{equation}
where $x = f(u_1,u_2)$ is a deterministic function. \\

Recently polar codes have also been applied to the classical broadcast channel \cite{GAG13, MHSU14}. 
In \cite{MHSU14} the authors show how polar codes can be used for the general broadcast channel to achieve the Marton-Gelfand-Pinsker region with and without common messages. 

In the remainder of this chapter we will show that we can use the approach of \cite{MHSU14} to also achieve the Marton-Gelfand-Pinsker region with and without common messages for classical-quantum broadcast channels, giving rise to the largest rate region for the classical-quantum broadcast channel.

\subsection{Marton-Gelfand-Pinsker region for private messages}\label{MGPpm}

In this section we will show how to achieve the Marton-Gelfand-Pinsker region without common messages for classical-quantum broadcast channels. 
We will use the technique of alignment as we did for universal polar codes in Section \ref{cqUChannel} to achieve the rate region
\begin{align}
\begin{split}\label{marton}
R_1 &\leq I(V,V_1;B_1), \\
R_2 &\leq I(V,V_2;B_2) \\
R_1 + R_2 &\leq I(V,V_1;B_1) + I(V_2;B_2|V) - I(V_1;V_2|V), \\
R_1 + R_2 &\leq I(V,V_2;B_2) + I(V_1;B_1|V) - I(V_1;V_2|V).
\end{split}
\end{align}
for a classical-quantum two-user broadcast channel described by a classical input $X = \varphi(V, V_1, V_2)$ and a quantum output $\rho_x^{B_1B_2}$. 

Let $V, V_1, V_2$ be auxiliary binary random variables with $(V, V_1, V_2) \sim p_V p_{V_2|V} p_{V_1|V_2V}$. Now, let $X = \varphi(V, V_1, V_2)$ be a deterministic function. 
Without loss of generality we look at a broadcast channel such that $I(V; B_1) \leq I(V;B_2)$. 

Set
\begin{align}
U_{(0)}^{n} = V^{n}G_n, \\
U_{(1)}^{n} = V_1^{n}G_n, \\
U_{(2)}^{n} = V_2^{n}G_n,
\end{align}
with $G_n$ being the usual polar coding transformation. 

The variable $U_{(1)}^{n}$ carries the message of the first user, while $U_{(0)}^{n}$ and $U_{(2)}^{n}$ carry the message of the second user. 

From a coding point of view $V$ comes from superposition because $U_{(0)}^{n}$ can be decoded by both receivers but carries information only for the second one and $V_1, V_2$ come from the binning scheme and can only be decoded by one receiver respectively.  

Since we have to cope with these additional auxiliary variables we use the polarization technique introduced in Section \ref{asy} for achieving the asymmetric capacity of a channel. Hence we introduce sets to determine the polarization of the probability distribution for the input and the channel independently. For the case without auxiliary variables this reduces to the usual channel coding described in Section \ref{scq}. 

Define for $(l\in\{ 1,2\})$, with $Z(X\mid B)$ defined in Equation \ref{Z} in the Preliminaries, the following sets, which are explained below, 
\begin{align}
\begin{aligned}
\HH_{V} &= \{ i\in [n] : Z(U_{(0),i}\mid U_{(0)}^{n-1} ) \geq \delta_n \}, \\
\LL_{V}  &= \{ i\in [n] : Z(U_{(0),i}\mid U_{(0)}^{n-1} ) \leq \delta_n \},  \\
\HH_{V|B_l} &= \{ i\in [n] : Z(U_{(0),i}\mid U_{(0)}^{n-1} B_{(l)}^{n} ) \geq \delta_n \}, \\
\LL_{V|B_l}  &= \{ i\in [n] : Z(U_{(0),i}\mid U_{(0)}^{n-1} B_{(l)}^{n} ) \leq \delta_n \}, \\
\HH_{V_l|V} &= \{ i\in [n] : Z(U_{(l),i}\mid U_{(l)}^{n-1} U_{(0)}^{n} ) \geq \delta_n \}, \\
\LL_{V_l|V}  &=  \{ i\in [n] : Z(U_{(l),i}\mid U_{(l)}^{n-1} U_{(0)}^{n}  ) \leq \delta_n \}, \\
\HH_{V_l|V,B_l} &= \{ i\in [n] : Z(U_{(l),i}\mid U_{(l)}^{n-1} U_{(0)}^{n} B_{(l)}^{n} ) \geq \delta_n \}, \\
\LL_{V_l|V,B_l}  &=  \{ i\in [n] : Z(U_{(l),i}\mid U_{(l)}^{n-1} U_{(0)}^{n} B_{(l)}^{n}  ) \leq \delta_n \}, \\
\HH_{V_1|V,V_2} &= \{ i\in [n] : Z(U_{(1),i}\mid U_{(1)}^{n-1} U_{(0)}^{n} U_{(2)}^{n} ) \geq \delta_n \}, \\
\LL_{V_1|V,V_2}  &=  \{ i\in [n] : Z(U_{(1),i}\mid U_{(1)}^{n-1} U_{(0)}^{n} U_{(2)}^{n}  ) \leq \delta_n \},
\end{aligned}
\end{align}
which due to the polarization effect satisfy
\begin{align}
\begin{aligned}
\lim_{n\rightarrow\infty} \frac{1}{n} |\HH_{V}| &= H(V), &\lim_{n\rightarrow\infty} \frac{1}{n} |\LL_{V}| &= 1 - H(V), \\
\lim_{n\rightarrow\infty} \frac{1}{n} |\HH_{V|B_l}| &= H(V|B_l), &\lim_{n\rightarrow\infty} \frac{1}{n} |\LL_{V|B_l}| &= 1 - H(V|B_l), \\
\lim_{n\rightarrow\infty} \frac{1}{n} |\HH_{V_l|V}| &= H(V_l|V), &\lim_{n\rightarrow\infty} \frac{1}{n} |\LL_{V_l|V}| &= 1 - H(V_l|V), \\
\lim_{n\rightarrow\infty} \frac{1}{n} |\HH_{V_l|V,B_l}| &= H(V_l|V,B_l), &\lim_{n\rightarrow\infty} \frac{1}{n} |\LL_{V_l|V,B_l}| &= 1 - H(V_l|V,B_l), \\
\lim_{n\rightarrow\infty} \frac{1}{n} |\HH_{V_1|V,V_2}| &= H(V_1|V,V_2), &\lim_{n\rightarrow\infty} \frac{1}{n} |\LL_{V_1|V,V_2}| &= 1 - H(V_1|V,V_2), 
\end{aligned}
\end{align}
The polarization of the classical quantities follows from polar codes for classical source coding \cite{A10}, while the polarization of the quantities with quantum side information is shown in \cite{SRDR13}. 

Intuitively $\HH_{V}$ and $\LL_{V}$ describe the polarization of the random variable $V$, that means whether the $i$th bit is nearly completely deterministic given by the previous bits or not. 

Similarly $\HH_{V|B_l}$ and $\LL_{V|B_l}$ denote whether the $l$th receiver can decode bits knowing the previous inputs and all outputs. 

$\HH_{V_l|V}$, $\LL_{V_l|V}$, $\HH_{V_l|V,B_l}$, $\LL_{V_l|V,B_l}$ have the same interpretation just with the additional side information from decoding $V$ first. 

$\HH_{V_1|V,V_2}$ and $\LL_{V_1|V,V_2}$ handles the indicies decoded by the first user with assumed knowledge of $V$ and $V_2$, while in our case he does not have access to the latter.

To easily see that this reduces to channel coding consider $\HH_{V|B_1}$ which gives a rate $1 - H(V|B_1)$. Assuming a uniformly distributed input this becomes
\begin{align}
\begin{aligned}
&1 - H(V|B_1) \\
&= H(V) - H(V|B_1) \\
&= H(V) + H(B_1) - H(VB_1) \\
&= I(V;B_1), 
\end{aligned}
\end{align}
since in this case $H(V)=1$. \\

First we look at $U_{(0)}^{n}$ which can be decoded by both users but only contains information for the second one. Define $\II^{(2)}_{sup} = \HH_V \cap \LL_{V|B_2}$ containing positions decodable for the second user and $\II_v^{(1)} = \HH_V \cap \LL_{V|B_1}$ containing positions decodable for the first user. 

$U_{(2)}^{n}$ can only be decoded by the second user and also only contains information for this receiver, by $\II^{(2)}_{bin} = \HH_{V_2|V} \cap \LL_{V|B_2}$ we denote the set of indices which can be decoded by the second receiver. 

$U_{(1)}^{n}$ only needs to be decoded by the first user and also only contains information for that user. $\II^{(1)} = \HH_{V_1|V} \cap \LL_{V|B_1}$ denotes the set of indices which the first receiver can decode reliably. 

We also have to take into account that the first user cannot decode $U_{(2)}^{n}$ therefore the indices in the set $\FF^{(1)} = \LL_{V_1|V,V_2} \cap \HH_{V_1|V} \cap \HH_{V_1|V,B_1}$ are critical. \\

We now use chaining constructions like those for universal polar codes in Section \ref{cqUChannel}. 
In total we need three different steps of alignment. 

First we handle $U_{(0)}^{n}$, by definition those should be decoded by both receivers and contain information only for the second one. 
We simply use the alignment technique for classical-quantum channels to send the message assigned for the second receiver to both of them. By the assumption that $I(V; B_1) \leq I(V;B_2)$ we get that we can reliably send information of amount $I(V; B_1)$ to both users. These sets contain information for the second user. We know that whenever $I(V; B_1)$ is not equal to $I(V; B_2)$ there will be unaligned indices left. Lets call this set $\BB^{(2)}$. 

In the second step we choose a subset $\BB^{(1)}$ of $\II^{(1)}$ such that $|\BB^{(1)}| = |\BB^{(2)}|$. We can then align these two subsets and therefore raise the number of indices from $U_{(0)}^{n}$ which both receivers can decode to $I(V; B_2)$. 

In the third step we need to cope with the fact that the first user can not decode the informations in $\FF^{(1)}$. 
Again choose a subset $\RRR_{bin}$ of $\II^{(1)}$ such that $|\RRR_{bin}| = |\FF^{(1)}|$. We use $\RRR_{bin}$ to repeat the information for the first user in $\FF^{(1)}$ of the following block. 

In order to get the correct order for the successive cancellation decoder we will encode $U_{(0)}^{n}$ and $U_{(2)}^{n}$ forward, while $U_{(1)}^{n}$ will be decoded backwards. 
Moreover, the first receiver decodes $U_{(0)}^{n}$ and $U_{(1)}^{n}$ forwards, while the second receiver decodes $U_{(0)}^{n}$ and $U_{(2)}^{n}$ backwards. 
\pgfooclass{ssstamp}{ 
    \method ssstamp() { 
    }
 \method cnot(#1,#2,#3,#4) { 
	\draw (#1,#2) -- (#1-#3,#2) -- (#1-#3,#2-#4) -- (#1,#2-#4);
	\draw (#1-#3,#2) circle (0.1);
	\draw (#1-#3,#2+0.1) -- (#1-#3,#2) -- (#1-#3-0.25,#2);
	\draw[dotted] (#1-#3-0.25,#2) -- (#1-#3-0.75,#2);
   }
 \method box(#1,#2,#3) { 
	\draw (#1,#2) -- (#1+4,#2) -- (#1+4,#2+4) -- (#1,#2+4) -- cycle; 
   }
}
\pgfoonew \myssstamp=new ssstamp()
\begin{figure}[t]
\centering
\begin{tikzpicture}[scale=0.8]
	\node[blue] at (0+2,10+0.5) {$\II_v^{(1)}$};
	\node[red] at (0+2,10+3.3) {$\II^{(2)}_{sup}$};
	\node[blue] at (5+2,10+0.5) {$\II_v^{(1)}$};
	\node[red] at (5+2,10+3.3) {$\II^{(2)}_{sup}$};
	\node[blue] at (10+2,10+0.5) {$\II_v^{(1)}$};
	\node[red] at (10+2,10+3.3) {$\II^{(2)}_{sup}$};
	\draw[line width=2pt, blue] (0,10) -- (0+4,10+4) -- (0+4,10) -- cycle; 
	\draw[line width=2pt, blue] (5,10) -- (5+4,10+4) -- (5+4,10) -- cycle; 
	\draw[line width=2pt, blue] (10,10) -- (10+4,10+4) -- (10+4,10) -- cycle; 
	\draw[line width=2pt, red] (0,14) -- (0+4,10+4) -- (0+4,10) -- cycle; 
	\draw[line width=2pt, red] (5,14) -- (5+4,10+4) -- (5+4,10) -- cycle; 
	\draw[line width=2pt, red] (10,14) -- (10+4,10+4) -- (10+4,10) -- cycle; 
	\node[red] at (0+2,0+0.5) {$\II^{(2)}_{bin}$};
	\node[red] at (5+2,0+0.5) {$\II^{(2)}_{bin}$};
	\node[red] at (10+2,0+0.5) {$\II^{(2)}_{bin}$};
	\draw[line width=2pt, red] (0,0) -- (0+4,0+0) -- (0+4,2) -- (0,2) -- cycle; 
	\draw[line width=2pt, red] (5,0) -- (5+4,0+0) -- (5+4,2) -- (5,2) -- cycle; 
	\draw[line width=2pt, red] (10,0) -- (10+4,0+0) -- (10+4,2) -- (10,2) -- cycle; 
	\draw[line width=2pt, blue] (0,5) -- (0+4,5+0) -- (0+4,5+2) -- (0,5+2) -- cycle; 
	\draw[line width=2pt, blue] (5,5) -- (5+4,5+0) -- (5+4,5+2) -- (5,5+2) -- cycle; 
	\draw[line width=2pt, blue] (10,5) -- (10+4,5+0) -- (10+4,5+2) -- (10,5+2) -- cycle; 
	\myssstamp.box(0,0,2);
	\myssstamp.box(5,0,2);
	\myssstamp.box(10,0,2);
	\myssstamp.box(0,5,1);
	\myssstamp.box(5,5,1);
	\myssstamp.box(10,5,1);
	\myssstamp.box(0,10,0);
	\myssstamp.box(5,10,0);
	\myssstamp.box(10,10,0);
           \draw[->,solid] (2.5,10.5)  -- (6,13.5); 
           \draw[->,solid] (7.5,10.5)  -- (11,13.5); 
	\draw[densely dotted] (0.5,7.5) -- (1.5,7.5) -- (1.5,8.5) -- (0.5,8.5) -- cycle;
	\draw[densely dotted] (5.5,7.5) -- (6.5,7.5) -- (6.5,8.5) -- (5.5,8.5) -- cycle;
	\draw[densely dotted] (10.5,7.5) -- (11.5,7.5) -- (11.5,8.5) -- (10.5,8.5) -- cycle;
	\draw[densely dotted] (2.5,5.5) -- (3.5,5.5) -- (3.5,6.5) -- (2.5,6.5) -- cycle;
	\draw[densely dotted] (7.5,5.5) -- (8.5,5.5) -- (8.5,6.5) -- (7.5,6.5) -- cycle;
	\draw[densely dotted] (12.5,5.5) -- (13.5,5.5) -- (13.5,6.5) -- (12.5,6.5) -- cycle;
	\node at (1,8) {$\FF^{(1)}$};
	\node at (6,8) {$\FF^{(1)}$};
	\node at (11,8) {$\FF^{(1)}$};
	\node at (3,6) {$\RRR_{bin}$};
	\node at (8,6) {$\RRR_{bin}$};
	\node at (13,6) {$\RRR_{bin}$};
           \draw[->,solid] (11.6,12.6)  -- (7.15,6.25); 
           \draw[->,solid] (6.6,12.6)  -- (2.15,6.25); 
           \draw[->,solid] (10.75,7.75)  -- (8.25,6.25); 
           \draw[->,solid] (5.75,7.75)  -- (3.25,6.25); 
	\draw[densely dotted] (1.3,5.5) -- (2.3,5.5) -- (2.3,6.5) -- (1.3,6.5) -- cycle;
	\draw[densely dotted] (6.3,5.5) -- (7.3,5.5) -- (7.3,6.5) -- (6.3,6.5) -- cycle;
	\draw[densely dotted] (11.3,5.5) -- (12.3,5.5) -- (12.3,6.5) -- (11.3,6.5) -- cycle;
	\node at (1.8,6) {$\BB^{(1)}$};
	\node at (6.8,6) {$\BB^{(1)}$};
	\node at (11.8,6) {$\BB^{(1)}$};
	\draw[densely dotted] (1.0,13) -- (3,13);
	\draw[densely dotted] (6.0,13) -- (8,13);
	\draw[densely dotted] (11.0,13) -- (13,13);
	\node at (2,12.7) {$\BB^{(2)}$};
	\node at (7,12.7) {$\BB^{(2)}$};
	\node at (12,12.7) {$\BB^{(2)}$};
	\node[blue] at (0.7,6) {$\II^{(1)}$};
	\node[blue] at (5.7,6) {$\II^{(1)}$};
	\node[blue] at (10.7,6) {$\II^{(1)}$};
	\node at (-1,12) {$U_{(0)}$};
	\node at (-1,7) {$U_{(1)}$};
	\node at (-1,2) {$U_{(2)}$};
\end{tikzpicture}
\caption[Coding for the broadcast channel.]{Coding for the broadcast channel. Indices in blue subsets are good for the first receiver and those in red subsets are good for the second receiver. Arrows indicate the alignment process.}
\end{figure}
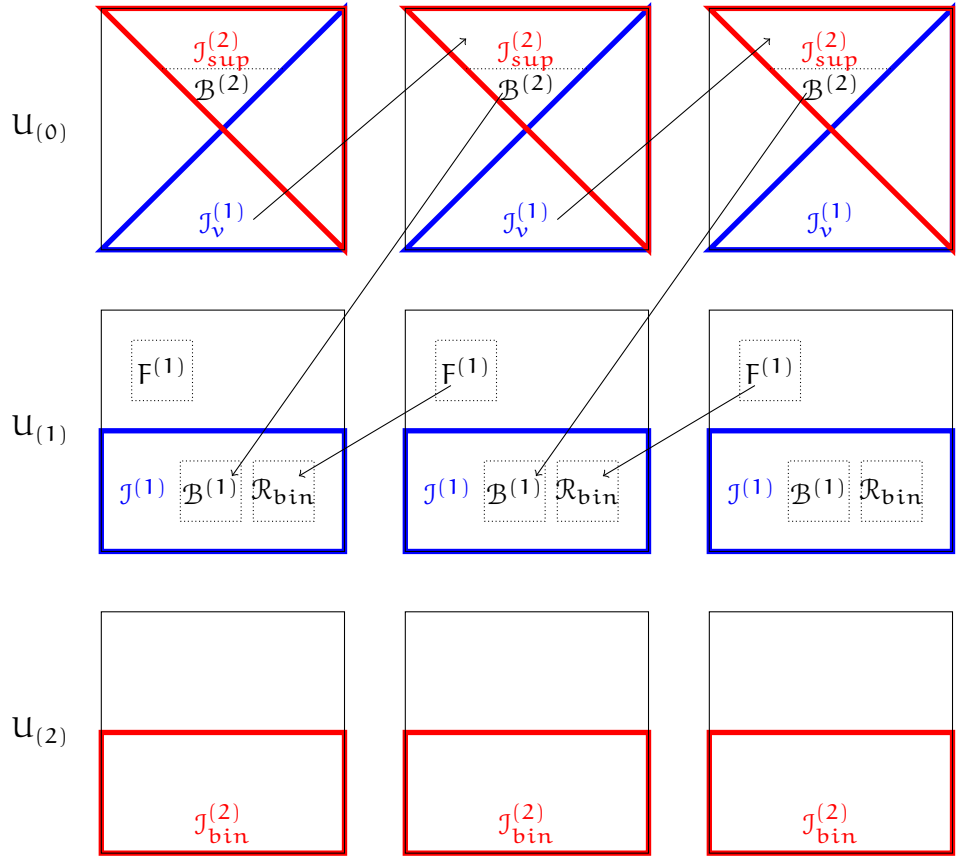

Now if we let the number of blocks approach infinity we can calculate the rate for the first receiver as 
\begin{align}
R_1 &= \frac{1}{n} (|\II^{(1)}| - |\BB^{(1)}| - |\RRR_{bin}|) \\
&= I(V_1; B_1 \mid V) - I(V_1; V_2\mid V) - (I(V;B_2) - I(V;B_2)) \\
&= I(V,V_1; B_1) - I(V_1; V_2\mid V) - I(V;B_2). 
\end{align}
We can also calculate the rate for the second receiver
\begin{align}
R_2 &= \frac{1}{n}(|\II^{(2)}_{sup}| + |\II^{(2)}_{bin}|) \\
&= I(V; B_2) + I(V_2; B_2\mid V) \\
&= I(V,V_2;B_2). 
\end{align}

Finally note that if we swap the role of the two receivers, the set $\BB^{(2)}$ will be empty due to the assumption that $I(V; B_1) \leq I(V;B_2)$, therefore we can achieve the rates
\begin{align}
R_1 &= I(V,V_1; B_1) \\
R_2 &= I(V_2; B_2\mid V) - I(V_1; V_2\mid V)
\end{align}

For the classical case it is known \cite{MHSU14} that these two rate pairs coincide with the Marton-Gelfand-Pinsker rate region. The proof can be directly translated to the setting of classical quantum communication and therefore our scheme achieves the rate region stated in \ref{marton}.

\subsection{Marton-Gelfand-Pinsker region with common messages}

We can simply extend our coding scheme in Section \ref{MGPpm} to a setting with a common message for both receivers by noting that the information sent via $U_{(0)}^{n}$ can be reliably decoded by both users. Therefore we can use these indices to send an amount of information equal to $\min\{ I(V;B_1), I(V;B_2)\}$ to both users. 
This leads to the rate region
\begin{align}
\begin{split}\label{macm}
R_0  &\leq \min\{ I(V;B_1), I(V;B_2)\}, \\
R_0 + R_1 &\leq I(V,V_1;B_1), \\
R_0 + R_2 &\leq I(V,V_2;B_2) \\
R_0 + R_1 + R_2 &\leq I(V,V_1;B_1) + I(V_2;B_2|V) - I(V_1;V_2|V), \\
R_0 + R_1 + R_2 &\leq I(V,V_2;B_2) + I(V_1;B_1|V) - I(V_1;V_2|V).
\end{split}
\end{align}

\newpage
\chapter{Quantum Channel}\label{qqc}
\section{Polar Codes for quantum channels}\label{qChannel}
In this section we will discuss how polar codes can be applied to quantum channels. 
We can roughly sort the existing schemes into two approaches, which here we will call "simultaneous" and "concatenated" scheme. 
Both have in common that polar codes are used to communicate phase and amplitude informsation reliably over a channel and thus get a reliable quantum code. This is based on an idea in \cite{RB08}, that is, transmitting amplitude and phase information will allow us to share maximally entangled bits between sender and receiver. 

In the simultaneous schemes amplitude and phase are polarized together \cite{WG13QDeg, RDR12, WR12}, while in the concatenated schemes amplitude and phase are polarized at different stages of the protocol \cite{SRDR13}. 

In the remainder of this section we present one example for each, the simultaneous and the concatenated scheme. 

\subsection{Simultaneous scheme}
Here we review the scheme introduced in \cite{WR12} for quantum communication over arbitrary channels. 

The first step is to define the amplitude and phase channels induced by a given quantum channel $\NN$. 
The amplitude channel is a cq-channel which sends an amplitude basis state over $\NN$ defined as
\begin{equation}\label{ac}
W_A : z \rightarrow \NN^{A'\rightarrow B} (\ketbra{}{z}{z}) \equiv \varphi_z^B,
\end{equation}
the phase channel is similarly defined as a cq-channel sending a state modulated in phase basis but with amplitude as quantum side information
\begin{equation}\label{pc}
W_P : x \rightarrow \sigma_x^{BC}, 
\end{equation}
where the output state is defined by 
\begin{equation}
\ket{}{\sigma_x}^{BCR} = U_{\NN}^{A'\rightarrow BR} (Z^x)^C \ket{}{\Phi}^{A'C},
\end{equation}
where $Z$ is the Pauli operator used to modulate $x$ into the signal by rephasing the amplitude basis state and $\ket{}{\Phi}^{A'C}$ is the maximally entangled state on $A'C$. 
Here $U_{\NN}^{A'\rightarrow BR}$ denotes the isometric extension of $\NN$ (see Equation \ref{stine}) and the system $C$ describes the quantum side information. This side information is sent to Bob over a noiseless channel and can be seen as the information available to Bob after decoding the amplitude channel given by possible correlations between amplitude and phase channel. 

The encoding is equivalent to the original polar coding scheme in \cite{A09} (or similarly to the classical-quantum coding described in Section \ref{cq}) exchanging classical CNOT gates by quantum CNOT gates. 

By the results in Section \ref{cq} it follows that the polarization induces synthesized amplitude channels  $W_{A,N}^{(i)}$ and that the fraction of those channels which are good is equal to the symmetric capacity of the induced amplitude channel
\begin{equation}
I(W_A) = I(Z^A;B). 
\end{equation}   
As a result from~\cite{RDR12} we also know that the phase channel $W_P$ polarizes, due to the side information with the fraction of good channels being equal to 
\begin{equation}
I(W_P) = I(X^A;BC). 
\end{equation}
An interesting fact to note is that the phase channel is polarized by the transpose of the usual transformation, namely $G_N^T$. Hence for the synthesized phase channels $W_{P,N}^{(i)}$ all future bits are known while the previous are not, the reverse of the usual case for polar codes. 

Similar to Equation \ref{sets} in Section \ref{comMAC} we can define sets including the indices of the good or bad synthesized channels for amplitude and phase respectively. By $G_N(W_A, \beta)$ we denote the set of channels good for amplitude and by $B_N(W_A, \beta)$ the set of channels bad for amplitude. Equivalently $G_N(W_P, \beta)$ and $B_N(W_P, \beta)$ for the phase channel. 
Considering these four sets we can divide the inputs to the quantum channel $\NN$ into the following sets: 
\begin{align}
A_N \equiv G_N(W_A, \beta) \cap G_N(W_P, \beta), \\
X_N \equiv G_N(W_A, \beta) \cap B_N(W_P, \beta), \\
Z_N \equiv B_N(W_A, \beta) \cap G_N(W_P, \beta), \\
B_N \equiv B_N(W_A, \beta) \cap B_N(W_P, \beta).
\end{align}
The set $A_N$ denotes the inputs which are good in amplitude and phase and can thus be used to send information bits. The sets $X_N$ and $Z_N$ denote inputs which are only good in amplitude or phase therefore we send bits frozen in phase or amplitude basis respectively. Finally inputs in the set $B_N$ are bad in amplitude and phase and hence we need to send halves of ebits. 
Thus the resulting code is entanglement assisted. 

It follows that the rate of the scheme is given by 
\begin{equation}
R_Q(\NN) = \frac{|A_N|-|B_N|}{N}.
\end{equation}
It is proven in \cite{WR12} that the rate is equal to
\begin{equation}
R_Q(\NN) = I(A\rangle B)_{\NN}
\end{equation}
and therefore the symmetric capacity $I_{sym}(\NN)$ of the channel. 

Also its proven in \cite{WR12} that for large enough $N$ the quantum polar coding scheme has an error upper bounded by $o(2^{-\frac{1}{2}N^{\beta}})$, for $\beta < \frac{1}{2}$. 
Again the encoding can be done efficiently, while this is not clear for the decoding. 

As described above the scheme is entanglement assisted, but \cite{WR12} provides two special cases where the fraction of entanglement needed vanishes.  
First this is shown for degradable channels, by relating the properties of the phase channel with induced amplitude channel from the sender to a reference system. 
Then it is shown that the amount of entanglement needed vanishes when $F(W_A) + F(W_P) \leq 1$ by bounding the output fidelity of the amplitude and phase channel by that of a binary erasure channel. 

\subsection{Concatenated scheme}

In this section we briefly review the concatenated scheme introduced in \cite{SRDR13}. Moreover we will focus on its application to polar codes rather than describing the general approach for CSS codes described in \cite{SRDR13}. 

In this scheme polarization is applied in two stages, namely via the so called \emph{outer} and \emph{inner} encoders. 

First the input states are polarized in phase basis at the outer encoder, therefore frozen qubits are added which are random in phase basis. The encoding can be written as
\begin{equation}
V^M = \sum_{x\in\{ 0,1\}^{KM}} \ketbra{}{(\tilde G^K_M)^T \tilde x}{\tilde x},
\end{equation}
where $\tilde G^K_M$ denotes $K$ parallel applications of the usual polar coding transformation $G_M$.
This leads to a multilevel coding scheme in order to assure the polarization at the inner decoder. 

Then at the inner encoder we add frozen qubits random in amplitude basis and also polarize in amplitude basis. 
For the inner layer inputs are actually polarized with $M$ polar coding blocks of length $L$, giving a block length of the overall scheme of $N=ML$. For each block the encoder can be written as
\begin{equation}
\bar V^L = \sum_{z^L\in\{ 0,1\}^{L}} \ketbra{}{G_L z^L}{z^L}.
\end{equation}

Thus it is required to code for \emph{sequences} of bits instead of single bits at the outer decoder, therefore the above mentioned multilevel coding scheme is required (for details we refer to \cite{SRDR13}). 

Note that we only add frozen bits random in amplitude or phase basis, thus this scheme is not entanglement assisted. \\

It is further shown that the described scheme achieves the coherent information 
\begin{equation}
R_Q(\NN) = I(A\rangle B)_{\NN},
\end{equation}
with an error, determined by the trace distance between the decoded and the ideal state, less than
\begin{equation}
\sqrt{2\epsilon_2} + \sqrt{2M\epsilon_1},
\end{equation}
where 
\begin{align}
\begin{aligned}
\epsilon_1 &= O(2^{-L^{\beta}}) \\
\epsilon_2 &=  O(L 2^{-M^{\beta'}} + L 2^{-\frac{1}{2}M^{\beta''}})
\end{aligned}
\end{align}
for $\beta, \beta', \beta'' <\frac{1}{2}$.

\newpage
\section{Towards universal polar codes for quantum channel}\label{UQC}
Similar to the classical-quantum case we are interested in transmitting \emph{quantum} information over a quantum channel when we don't know the exact channel but only that its in a predefined set of channels.
Naturally one would expect that we can use the schemes described in Section \ref{qChannel} and use techniques similar to those for classical and classical-quantum channels (see Section \ref{cqUChannel}) to achieve universal polar codes for quantum channels. As it turns out some difficulties arise from simultaneous handling of the amplitude and phase information, when attempting to get a reliable coding for the compound quantum channel. 

In these sections we will describe possible approaches to universal quantum polar codes achieving the compound capacity in certain special cases. 

\subsection{Universal polar codes for degraded channels in the simultaneous scheme}
We call a channel $\NN_1$ degraded with respect to another channel $\NN_2$, denoted by $\NN_1 \preceq \NN_2$, if there exists a degrading channel $\DD$ such that $\NN_1 = \DD (\NN_2)$. 

In the case of classical and classical-quantum channels it is known that the so-called ``subset property'' holds, that is, whenever a classical or classical-quantum channel $W_1$ is degraded with respect to another channel $W_2$ then the set of indices which are good for the channel $W_1$ is a subset of the set of indices which are good for the channel $W_2$. This can easily be seen by considering the data processing inequality for the mutual information (see Equation \ref{dpi}). 

Therefore if these two channels form a compound channel, the compound capacity can always be achieved by coding for $W_1$ because the good indices for this channel are automatically good for both channels. Indeed no alignment is needed in this special case. 

In this section we will show that a similar statement holds also for polar codes for quantum channels using the simultaneous scheme. \\

We consider the compound channel $\NN$ consisting of the set $\{\NN_1, \NN_2\}$, where $\NN_1\preceq\NN_2$. 

First we note from the definitions of the phase and amplitude channel in Equations \ref{ac} and \ref{pc} that the induced channels $W_{A_1}$, $W_{P_1}$, $W_{A_2}$ and $W_{P_2}$ obey the following relations
\begin{align}
W_{A_1} \preceq W_{A_2}, \\
W_{P_1} \preceq W_{P_2}.
\end{align}
This follows from applying the degrading map $\DD$, defined by $\NN_1 = \DD (\NN_2)$ to the output states of the amplitude and phase channels $W_{A_2}$ and $W_{P_2}$.

Therefore the number of possible combinations corresponding to whether an index is good or bad for the first or second channel respectively reduces significantly by considering the subset property. 
Indeed we have to code for essentially four types of combinations.
\begin{table}[t]
\centering
\begin{tabular}
[c]{c|c|c|c|c}\hline\hline
 A1 & P1 & A2 & P2 & Coding \\\hline\hline
 G & G & G & G & $\ket{}{info}$ \\\hline
 G & B & G & G & $\ket{}{+}$\\\hline
 G & B & G & B & $\ket{}{+}$\\\hline
 B & G & G & G& $\ket{}{0}$\\\hline
 B & G & B & G& $\ket{}{0}$\\\hline
 B & B & G & G& $\ket{}{\Phi}$\\\hline
 B & B & G & B& $\ket{}{\Phi}$\\\hline
 B & B & B & G& $\ket{}{\Phi}$\\\hline
 B & B & B & B& $\ket{}{\Phi}$\\\hline\hline
\end{tabular}
   \caption{Remaining combinations for degradable channels}
   \label{tab:degChannel}
\end{table}

First the channels which are doubly good for $\NN_1$, these are a subset of the doubly good ones for $\NN_2$ so we can always simply send information bits. 

With the inverse argument the channels which are doubly bad for $\NN_2$ are a subset of the doubly bad ones for $\NN_1$, therefore we can just send halves of entangled states whenever a channel is doubly bad for $\NN_1$. 

The remaining two types are those which we can simply freeze in either amplitude or phase basis for $\NN_1$, again because these channels must be at least good in phase or amplitude basis respectively for $\NN_2$. 

Hence we can code at a rate $I(A\rangle B)_{\NN_1}$ and therefore send information at the capacity rate of the compound channel. \\

It can easily be seen that the above construction also works for larger sets consisting of k channels with
\begin{equation}
\NN_1\preceq\NN_2\preceq\dots\preceq\NN_k
\end{equation}
for coding at a rate $I(A\rangle B)_{\NN_1}$.

\newpage
\chapter{Conclusions and outlook}\label{CO}
\section{Further directions} \label{fd}

In this section we would like to emphasize some possible directions for further investigations of polar codes in quantum information theory. \\

First, from the results in Chapter \ref{cqc} we would conjecture that it is possible to generalize many communication models from a fully classical scenario to classical-quantum settings. In general this should be possible for all channels where the classical achievability can be shown using the same tools as the channels discussed in this work. 

In particular it would be interesting to apply polar codes to channels such as the classical-quantum wiretap channel, where we consider an additional eavesdropper modeled by a channel from the sender to an additional system, because this would require investigating complementary constraints, here a secrecy criterion. Polar codes for classical wiretap channels have been investigated in \cite{SV13, WU14}. \\

As described previously, classical polar codes have an efficient encoding and decoding, while for quantum settings we only know that the encoding can be done efficiently. One of the most important open questions is whether an efficient decoding is also possible for quantum channels, in order to apply polar codes to practical scenarios. Recent results in this direction are presented in \cite{WLH13}. \\

Also in the direction of practical implications one could investigate the achievable rates by polar codes using a limited number of channel uses. Usually we look at the effect of polarization with the number of channels uses going to infinity, it would be interesting to quantify the impact of using a finite number of channel instances. Some work in this direction is done for classical polar codes, a generalization to quantum polar codes would be desirable. In the classical setting this question is addressed in \cite{HKU09, GV14}. \\

Another direction would be to extend the results in Chapter \ref{qqc} involving quantum communication. In particular Section \ref{UQC} gives rise to the question of how polar codes for quantum channels can be applied to compound quantum channels, in order to achieve the best known rate. As described in Section \ref{UQC}, the main difficulty is to find a suitable way of aligning channels such that amplitude and phase information can be handled sufficiently. \\

An additional interesting question was raised in \cite{SRDR13}, that is, whether the concatenated scheme for quantum communication described in Section \ref{qChannel} could also achieve a rate higher then the coherent information, for example the regularized quantum capacity (see Equation \ref{qcapcity} ). \\

It would also be interesting to investigate generalizations of the quantum polar coding schemes to settings with many users, such as multiple access channels or broadcast channels. \\

Finally one could explore extensions to the basic polar coding scheme, such as the recently introduced \emph{branching mera codes} \cite{FP13, FP14}. This could lead to improved results concerning the finite length coding and error rate. 

\newpage
\section{Conclusion}\label{Conclusion}

In this work we explored applications of polar codes to quantum information theory. 

First we showed how to apply polar codes to classical-quantum channels, as well as several generalized scenarios to multi-party settings and compound channels. Most notably, we were able to show that the Han-Kobayashi rate region for the classical-quantum two-user interference channels can be achieved using a successive decoder, hence in particular without the need of a simultaneous decoder. 

In addition we extended previous results to also achieve the asymmetric capacity of a classical-quantum setting and improved the block error probability for classical-quantum polar codes. 

Moreover were able to show a new achievable rate region for the classical quantum broadcast channels and hence give an example where polar codes can be used for proving achievability. \\

Additionally we showed how to extend polar codes to the transmission of quantum information over quantum channels and investigated possible directions for universal polar codes for quantum channels. In particular we showed that polar codes can be used to code at the optimal rate for compound channels consisting of a set of degraded channels. \\

Finally we gave ideas for possible further directions for exploring polar codes in quantum information theory. 
Based on the results presented in this work we believe that the technique of polarization can be applied to many other communication settings and will continue to play an important role in quantum information theory. 
\newpage   
  
\manualmark
\markboth{\spacedlowsmallcaps{\bibname}}{\spacedlowsmallcaps{\bibname}} 
\refstepcounter{dummy}
\addtocontents{toc}{\protect\vspace{\beforebibskip}} 
\bibliographystyle{plainnat}
\bibliography{Bib}

\begin{appendix}

\end{appendix}   
\end{document}